\documentclass[11pt]{article}
\usepackage{amsthm,amsmath,latexsym,multibox,amssymb,amsfonts,amsbsy}
\usepackage{lscape,fancyhdr,array,stmaryrd,euscript,wrapfig}
\usepackage{cite,wrapfig,color}
\usepackage{graphicx}

\newcommand{\bep}{\begin{picture}}
\newcommand{\eep}{\end{picture}}

\newcounter{YoungHeight}\newcounter{YoungWidth}

\newcounter{Mul1}\newcounter{Mul2}\newcounter{Mul3}\newcounter{Mul4}
\newcounter{A1}\newcounter{A2}
\newcounter{B3}
\newcounter{C3}\newcounter{C4}

\newcounter{T0}\newcounter{T1}

\newcounter{R0}\newcounter{R1}

\newlength{\txtHShift}

\newlength{\txtWidth}

\newcommand{\HalfLength}[2]{\setcounter{Mul1}{#1}\setcounter{Mul2}{#1}\addtocounter{Mul1}{\value{Mul2}}\addtocounter{Mul1}{\value{Mul2}}%
\addtocounter{Mul1}{\value{Mul2}}\addtocounter{Mul1}{\value{Mul2}}\setcounter{#2}{\value{Mul1}}}

\newcommand{\Add}[3]{\setcounter{#1}{#2}\addtocounter{#1}{#3}}

\newcommand{\Length}[1]{#10}
\newcommand{\YoungScale}{}

\newcommand{\shiftedText}[2]{{\hspace{#1}{#2}}}
\newcommand{\calcHShift}[1]{\settowidth{\txtWidth}{#1}\setlength{\txtHShift}{-0.5\txtWidth}}
\newcommand{\calcHShiftA}[1]{\settowidth{\txtWidth}{#1}\setlength{\txtHShift}{-0.3\txtWidth}}

\newcommand{\TextTop}[3]{{\calcHShift{#1}\HalfLength{#2}{T0}\Add{T1}{\Length{#3}}{-7}\put(\value{T0},\value{T1}){\shiftedText{\txtHShift}{#1}}}}

\newcommand{\TextUnderA}[3]{{\HalfLength{#2}{T0}\calcHShift{#1}\put(\value{T0},-7){\shiftedText{\txtHShift}{$\scriptstyle #1$}}}}
\newcommand{\TextUnderB}[3]{{\HalfLength{#2}{T0}\calcHShiftA{#1}\put(\value{T0},-7){\shiftedText{\txtHShift}{$\scriptstyle #1$}}}}

\newcommand{\BlockA}[2]{{\YoungScale\bep(\Length{#1},\Length{#2}){\Add{A1}{#1}{1}\Add{A2}{#2}{1}}%
\multiput(0,0)(10,0){\value{A1}}{\line(0,1){\Length{#2}}}\multiput(0,0)(0,10){\value{A2}}{\line(1,0){\Length{#1}}}%
\setcounter{YoungHeight}{\Length{#2}}\setcounter{YoungWidth}{\Length{#1}}\eep}}

\newcommand{\BlockB}[4]{{\YoungScale\Add{B3}{\Length{#2}}{\Length{#4}}%
\bep(\Length{#1},\value{B3})\put(0,\Length{#4}){\BlockA{#1}{#2}}%
\put(0,0){\BlockA{#3}{#4}}\setcounter{YoungHeight}{\value{B3}}\setcounter{YoungWidth}{\Length{#1}}\eep}}

\newcommand{\Rect}[2]{\bep(\Length{#1},\Length{#2})\put(0,0){\line(1,0){\Length{#1}}}\put(0,0){\line(0,1){\Length{#2}}}%
\put(\Length{#1},\Length{#2}){\line(-1,0){\Length{#1}}}\put(\Length{#1},\Length{#2}){\line(0,-1){\Length{#2}}}\eep}

\newcommand{\RectT}[3]{\bep(\Length{#1},\Length{#2})\put(0,0){\line(1,0){\Length{#1}}}\put(0,0){\line(0,1){\Length{#2}}}%
\put(\Length{#1},\Length{#2}){\line(-1,0){\Length{#1}}}\put(\Length{#1},\Length{#2}){\line(0,-1){\Length{#2}}}#3{#1}{#2}\eep}

\newcommand{\RectBRow}[4]{{\bep(\Length{#1},20)\put(0,0){\RectT{#2}{1}{\TextTop{#4}}}%
\put(0,10){\RectT{#1}{1}{\TextTop{#3}}}\eep}}

\newcommand{\YoungA}{\BlockA{1}{1}}
\newcommand{\YoungB}{\BlockA{2}{1}}

\newcommand{\YoungAA}{\BlockA{1}{2}}
\newcommand{\YoungBA}{\BlockB{2}{1}{1}{1}}
\newcommand{\YoungBB}{\BlockA{2}{2}}

\newcommand{\YoungCC}{\BlockA{3}{2}}

\newcommand{\YoungAAAA}{\BlockA{1}{4}}

\newcommand{\BlockCcA}[2]{{\YoungScale\bep(\Length{#1},\Length{#2}){\Add{A1}{#1}{1}\Add{A2}{#2}{1}}%
\multiput(0,0)(10,0){\value{A1}}{\line(0,1){\Length{#2}}}\multiput(0,0)(0,10){\value{A2}}{\line(1,0){\Length{#1}}}%
\multiput(0,0)(10,0){#1}{\multiput(0,0)(0,10){#2}{\qbezier(2,2)(5,5)(8,8)\qbezier(2,8)(5,5)(8,2)}}\eep}}

\newcommand{\YoungCcA}{\BlockCcA{1}{1}}
\newcommand{\YoungCcB}{\BlockCcA{2}{1}}
\newcommand{\YoungCcC}{\BlockCcA{3}{1}}
\newcommand{\YoungCcD}{\BlockCcA{4}{1}}

\newcommand{\YoungCcAA}{\BlockCcA{1}{2}}

\newcommand{\YoungCcBB}{\BlockCcA{2}{2}}

\newcolumntype{x}[1]{%
>{\centering\hspace{0pt}}m{#1}}%
\newcolumntype{w}[1]{%
>{\raggedright\hspace{0pt}}m{#1}}%
\newcolumntype{z}[1]{%
>{\raggedleft\hspace{0pt}}m{#1}}%

\newcommand{\dSAdS}{{\ensuremath{(A)dS_d}\,}}
\newcommand{\AdS}{{\ensuremath{AdS_d}\,}}

\newcommand{\ads}{{\ensuremath{\mathfrak{so}(d-1,2)}}}

\newcommand{\lorentz}{{\ensuremath{\mathfrak{so}(d-1,1)}}}

\newcommand{\msv}{{\ensuremath{\mathfrak{so}(d-1)}}}

\newcommand{\pl}{\partial}
\newcommand{\fm}[1]{_{\mathbf{{#1}}}}
\newcommand{\be}{\begin{equation}}
\newcommand{\ee}{\end{equation}}

\newcommand{\Ya}[1]{{\ensuremath{[#1]}}}

\newcommand{\Yy}{\ensuremath{\mathbf{Y}}}

\newcommand{\Ds}{\ensuremath{\mathbf{A}}}

\newcommand{\DL}{{D}}

\newcommand{\DO}{{D_{0}}}
\newcommand{\Sigm}{{\ensuremath{\boldsymbol{\sigma_-}}}}

\newcommand{\fud}[2]{{^{#1}_{\phantom{#1}#2}}}
\newcommand{\fdu}[2]{{_{#1}^{\phantom{#1}#2}}}

\newcommand{\fudu}[3]{{^{#1\phantom{#2}#3}_{\phantom{#1}#2}}}

\newcommand{\TwoColumn}[4]{\parbox[c]{30pt}{\unitlength=0.40mm\Add{R1}{\Length{#1}}{15}\bep(30,\value{R1})(0,0){%
\put(5,10){\Rect{1}{#1}}\Add{R0}{\Length{#1}}{-\Length{#2}}%
\put(15,10){\put(0,\value{R0}){\Rect{1}{#2}}}
\put(5,10){\TextUnderA{#3}{1}{1}}%
\put(15,10){\put(0,\value{R0}){\TextUnderB{#4}{1}{1}}}%
}\eep}}
\newcommand{\OneColumn}[2]{\parbox[c]{20pt}{\unitlength=0.40mm\Add{R1}{\Length{#1}}{15}\bep(20,\value{R1})(0,0){%
\put(5,10){\Rect{1}{#1}}%
\put(5,10){\TextUnderA{#2}{1}{1}}%
}\eep}}

\newcommand{\smallpic}[1]{{\unitlength=0.2mm#1}}

\newcommand{\boldpic}[1]{{\linethickness{0.4mm}#1}}

\definecolor{rougef}{rgb}{0.56,0,0}
\definecolor{vertf}{rgb}{0,0.5,0}
\definecolor{bleuf}{rgb}{0,0,0.8}

\newcommand{\Cl}{\ensuremath{\mathcal{C}\ell^{\phantom{0}}_{d-1,2}}}
\newcommand{\Cle}{\ensuremath{\mathcal{C}\ell^{{0}}_{d-1,2}}}
\begin{document}
\renewcommand{\thefootnote}{\fnsymbol{footnote}}
\begin{flushright}
\vspace{1mm}
\end{flushright}

\vspace{1cm}

\begin{center}
{\bf \Large  Higher-spin algebras and cubic interactions for simple
mixed-symmetry fields in AdS spacetime}

\vspace{2cm}

\textsc{Nicolas Boulanger\footnote{Research Associate of the Fund for
Scientific Research-FNRS (Belgium); nicolas.boulanger@umons.ac.be}
and E.D. Skvortsov\footnote{skvortsov@lpi.ru}}

\vspace{2cm}

{\em${}^*$ Service de M\'ecanique et Gravitation, Universit\'e de Mons -- UMONS\\
20 Place du Parc, 7000 Mons (Belgium)}
\vspace*{.5cm}

{\em${}^\dag$ P.N.Lebedev Physical Institute, Leninsky prospect 53, 119991, Moscow (Russia)}

\vspace{1cm}

\end{center}

\vspace{0.5cm}
\begin{abstract}
Nonabelian Fradkin--Vasiliev cubic interactions for
dual-graviton-like gauge fields with gravity and themselves are
constructed in anti-de Sitter spacetime. The Young diagrams of gauge
potentials have shapes of ``tall-hooks'', i.e. two columns the
second of height one.

The underlying nonabelian algebra is a Clifford algebra with the
anti-de Sitter signature. We also discuss the universal enveloping
realization of higher-spin algebras, showing that there is a
one-parameter family of algebras compatible with unitarity, which is
reminiscent of $d=3$ deformed oscillators.
\end{abstract}
\newpage

\tableofcontents

\renewcommand{\thefootnote}{\arabic{footnote}}
\setcounter{footnote}{0}

\section{Introduction}\setcounter{equation}{0}

Since the pioneering works by Fradkin and Vasiliev
\cite{Fradkin:1987ks,Fradkin:1986qy}
where the gravitational interaction problem (as well as self-interactions) for
higher-spin gauge fields was solved at the first nontrivial order by going to a
four-dimensional (anti-)de Sitter $(A)dS_4$ background,
there has been a lot of attention devoted to the study of totally-symmetric gauge
fields around $\AdS$ background, culminating with Vasiliev's fully nonlinear and
consistent equations for totally symmetric gauge fields
\cite{Vasiliev:1990en,Vasiliev:1992av,Vasiliev:2003ev}.
These equations admit constantly-curved spacetimes as exact solutions,
where it is crucial that the curvature be nonvanishing.
For a review of the key mechanisms of higher-spin extensions of gravity, see
\cite{Bekaert:2010hw}, while various reviews on Vasiliev's equations can be found
in \cite{Vasiliev:2000rn,Vasiliev:2004qz,Bekaert:2005vh}.

\vspace*{.2cm}

Depending on one's taste, one may view higher-spin gauge theory as a
limit of string theory \cite{Gross:1988ue}, or string theory as a
broken phase of higher-spin gauge theory. See however recent works
\cite{Polyakov:2009pk, Polyakov:2010qs, Polyakov:2011sm} where
higher-spin fields were realized as certain vertex operators in
exotic pictures without taking any limits of superstring theory. It
is expected that, for a better understanding of both higher-spin
gauge theory and strings together with their interconnections, the
role played by the fields at high levels, whose Young symmetries are
neither totally symmetric (first Regge trajectory) nor totally
antisymmetric ($p\,$-forms), will be crucial. For a recent
discussion and results, see \cite{Sagnotti:2010at}. At present, free
mixed-symmetry gauge fields are very well understood off-shell, in
both flat and $\dSAdS$ backgrounds. For a non-exhaustive list of
recent works on quadratic action principles and equations for mixed-symmetry gauge
fields in constantly-curved backgrounds, see e.g.
\cite{Bekaert:2003az,deMedeiros:2003dc, deMedeiros:2003px, Alkalaev:2003qv,
Sagnotti:2003qa, Alkalaev:2005kw, Alkalaev:2006rw,
Bekaert:2006ix,Fotopoulos:2007nm,
Buchbinder:2007ix,Reshetnyak:2008gp,Skvortsov:2008sh,Zinoviev:2008ve,Campoleoni:2008jq,
Boulanger:2008up,Boulanger:2008kw,
Campoleoni:2009gs,Zinoviev:2009gh,Skvortsov:2009nv,Alkalaev:2009vm,
Skvortsov:2012at} and references therein.

\vspace*{.2cm}

As far as the problem of finding consistent interactions for
mixed-symmetry fields is concerned, some analysis have been done in
flat background
\cite{Fradkin:1991iy,Bekaert:2002uh,Boulanger:2004rx,Bekaert:2004dz,Metsaev:2005ar,Metsaev:2007rn},
but in $\dSAdS$ background almost nothing has been achieved apart
from the very recent works
\cite{Alkalaev:2010af,Zinoviev:2011fv,Boulanger:2011qt} (see also
the earlier works \cite{Sezgin:2001zs,Sezgin:2001yf}) and
\cite{Zinoviev:2010av}.

\begin{wrapfigure}{r}{2.0cm}
\bep(20,60)(-10,0)\put(0,10){\Rect{1}{4}}%
\put(0,0){\YoungA}\put(0,40){\YoungBA}%
\put(2,24){$k$}\eep
\end{wrapfigure}
It is the goal of the present paper to study the interaction problem
for the simplest class of mixed-symmetry gauge fields, \emph{i.e.}
those that are described in the metric-like formalism by potentials
of Young shape $[k,1]$, i.e. with two columns, the first of
arbitrary height (albeit constrained by the spacetime dimension
$d$), the second of height one. We call such fields ``tall hooks''.
In particular the graviton belongs to the spectrum as $[1,1]$ Young
shape. The spin-one Maxwell field is also present as degenerate
$[0,1]$ Young shape. A detailed analysis of the case $[2,1]$
corresponding to the dual graviton in five dimensions has been
provided very recently in \cite{Boulanger:2011qt} where special
emphasis was put on the St\"uckelberg formulation and the flat limit
that will not be considered here.

Higher-spin fields are to be organized in the adjoint multiplet of
some higher-spin algebra $\mathfrak{g}$. Any reasonable higher-spin
algebra must contain $\ads$ as a subalgebra\footnote{See
\cite{Fradkin:1986ka, Konshtein:1988yg, Vasiliev:2004cm} for more
detailed exposition and original results.}. Gauging of $\ads$ itself
describes gravity by virtue of the MacDowell-Mansouri-Stelle-West
\cite{MacDowell:1977jt, Stelle:1979aj} approach. In addition
$\mathfrak{g}$ contains some other generators corresponding to
higher-spin fields.

The higher-spin algebra whose gauging leads to the spectrum built of
tall hooks is just the Clifford algebra, $\Cl$, with the anti-de
Sitter signature. The graviton appears when gauging $\Cl$ since
$\ads\in\Cl\,$. The spin-one field corresponds to the unity of $\Cl$.
Higher products of $\gamma$-matrices represents generators for
genuine tall hooks. The Clifford algebra $\Cl$ can be thought of as the
simplest higher-spin algebra because it is finite-dimensional.

We show that the structure constants of $\Cl$ are compatible with
switching on cubic interactions of tall hooks and construct the
cubic action using the Fradkin-Vasiliev approach,
\cite{Fradkin:1987ks,Fradkin:1986qy}. Yang-Mills groups can be
easily activated too, leading to colored tall hooks.

Aimed at giving an invariant definition for general higher-spin
algebras we consider the quotients of the universal enveloping
algebra of $\ads$, $U(\ads)$. The restrictions by unitarity lead
naturally to a one-parameter family $hs(\nu)$ of higher-spin
algebras with the even Clifford algebra and the Vasiliev higher-spin
algebra \cite{Vasiliev:2003ev} belonging to it at
$\nu=\frac{d(d+1)}{8}$ and at the ``singleton point''
$\nu=-\frac{(d-1)(d-3)}{4}$, respectively.

The algebra $hs(\nu)$ can be thought of as a generalization to arbitrary
dimension of the $3d$ higher-spin algebra of
\cite{Prokushkin:1998bq}, which is based on the deformed oscillator
algebra \cite{Vasiliev:1989re}. Recently there has been a lot of
interest in $3d$ $hs(\nu)$ in the context of $AdS_3/CFT^2$, see
e.g. \cite{Henneaux:2010xg, Campoleoni:2010zq, Campoleoni:2011hg,
Gaberdiel:2011wb}.

\vspace*{.2cm}

The outline of the paper is as follows. In Section \ref{sec:two} we
recall some results about two-column gauge fields in $\dSAdS$
background. Section \ref{sec:FV} reviews the Fradkin--Vasiliev
construction for consistent cubic interactions around $(A)dS$
background. In Section \ref{sec:FVgravitational} we present nonabelian
gravitational interactions for tall hooks. The manifestly
$\AdS$-covariant formulation for cubic interactions of tall hooks is
given in Section \ref{sec:AdScov}, where an extension of the
previous interactions is proposed, whereby the self-interactions are
considered as well. We discuss the realization of higher-spin
algebras in terms of $U(\ads)$ and some other developments in
Section \ref{sec:TallHooksDiscusssion}, before we give the
conclusions in Section \ref{sec:Conclusion}.

\section{Two-column gauge fields}\setcounter{equation}{0}
\label{sec:two}

In this Section we briefly review some features pertaining to the kinematics of
irreducible gauge fields in $\AdS$ backgrounds.

\subsection{Notation and conventions}

Base manifold indices, or world indices, are denoted by Greek letters
$\mu,\nu,\ldots\,$, while Lorentz indices are denoted by lower-case Latin letters.
The Lorentz algebra $\mathfrak{so}(d-1,1)$ is associated with the metric
$\eta_{ab}=\,$diag$(-,+,\ldots,+)$ where the indices $a,b,\ldots$ run over the values
$0,1,\ldots,d-1\,$, while the anti-de Sitter algebra $\mathfrak{so}(d-1,2)$ is associated
with the metric $\eta_{AB}=\,$diag$(-,+,\ldots,+,-)$ where the indices $A,B,\ldots$ run over the
values $0,1,\ldots,d-1,d\,$.

Square brackets indicate total antisymmetrization with strength one,
\emph{i.e.}
\begin{equation}
[a_1a_2\ldots a_p]=\frac{1}{p!}\;
            \sum_{\sigma\in\mathfrak{S}_p}(-1)^{|\sigma|}\;a_{\sigma(1)}\ldots a_{\sigma(p)}\quad ,
\end{equation}
where $\mathfrak{S}_p$ is the group of permutations of $p$ elements and $|\sigma|\in\{0,1\}$ is the
signature of the permutation $\sigma\in\mathfrak{S}_p\,$. Similarly, curved brackets
$(a_1a_2\ldots a_p)$ denote total symmetrization with strength one.
A group of $p$ totally antisymmetrized Lorentz indices will be denoted
by $[a_1a_2\ldots a_p]=a[p]\,$, while $(a_1a_2\ldots a_p)=a(p)$ for totally symmetrized indices,
and similarly for world and $\AdS$ indices.
To further simplify the notation we will often use conventions whereby like
letters imply complete (anti)symmetrization with strength one, so that
for example $e^a\omega^{a[3]}=e^{[a}\omega^{aaa]}\,$.
If not explicitly specified, the context will make clear whether
we mean complete symmetrization or antisymmetrization.

The components of an irreducible $\mathfrak{gl}_d$ tensor whose symmetry type
consists of a Young
diagram with two columns, the first of length $p$ and the second of length $q\,$,
will be denoted $\varphi^{\mu[p], \nu[q]}\;$, $p\geqslant q\,$.
The Young symmetry described above is abbreviated by $\varphi \sim [p,q]\,$.
The components of a Lorentz tensor of type $[p,q]$ are denoted by
$\varphi^{a[p], b[q]}\,$,
\emph{idem} for an  $\mathfrak{so}(d-1,2)$-tensor $\varphi^{A[p], B[q]}\,$.
Note that, by abuse of notation, we do not consider (anti) self-duality constraints
on what we call Lorentz (or $\AdS$) tensors in $d=2n\,$,
so that a Lorentz tensor, in our conventions, only obeys over-(anti)symmetrization
and trace constraints.
The torsion-free, Lorentz-connection on the base manifold is denoted by the symbol $D\,$.
In flat background, $\left[ D_\mu,D_\nu\right] = 0\,$ while in $\AdS$, one has
$\left[ D_{\mu},D_{\nu}\right] = \frac{1}{2}\;R_{\mu\nu}^{\quad ab} M_{ab}$
where $M_{ab}$ are the generators of
$\mathfrak{so}(d-1,1)$ and the components of the background curvature two-form
$R_{\mu\nu}^{\quad ab} = -\lambda^2 \,(h_{\mu}^{a} h_{\nu}^{b}-h_{\mu}^{b} h_{\nu}^{a})$
in terms of the $\AdS$ background vielbein components $h_{\mu}^{a}\,$
and the parameter $\lambda^2$ related to the cosmological constant $\Lambda$ by
$-\lambda^2=\frac{\Lambda}{(d-1)(d-2)}\;$.
We denote $D_a=h^{\mu}_a\, D_{\mu}$ featuring the inverse vielbein, or tetrad, components
$h^{\mu}_a\,$.

\subsection{Short review on two-column gauge fields in \AdS}

In this section we are mainly concerned with mixed-symmetry fields of
type-$[p,q]$, but also review some other background material.
\paragraph{On-mass shell.}
Considering a massive type-$[p,q]$ tensor in $\AdS$ spacetime,
the complete set of on-mass-shell conditions is
\begin{align}
(\square + m^2) \phi^{a[p],b[q]} & \; = \; 0\quad, \label{OnMassShellA}\\
 D_a\phi^{a[p],b[q]} & \;=\;0\;=\; D_b \phi^{a[p],b[q]} \quad,\label{OnMassShellB}\\
\eta_{ab}\;\phi^{a[p],b[q]} & \;=\; 0 \quad.\label{OnMassShellC}
\end{align}
For generic values of $m^2\,$, the equations (\ref{OnMassShellA})-(\ref{OnMassShellC})
possesses no gauge symmetries and describe a massive type-$[p,q]$ field.
Instead of the parameter $m^2\,$, it is convenient to use the minimal eigenvalue
$E_0$ (the lowest energy) of the $\mathfrak{so}(2)$ generator of the maximal compact subalgebra
$\mathfrak{so}(2)\oplus\mathfrak{so}(d-1)\subset$  \ads.
The lowest energy $E_0$ is related to $m^2$ by \cite{Metsaev:1995re}
\begin{equation}
m^2=\lambda^2\left[E_0(E_0-d+1)-p-q\right]\quad .
\label{WEMassFormula}
\end{equation}
There are three special values of $E_0$ at which certain gauge symmetry appears.
At each critical point the corresponding field equations modulo gauge transformations define an
irreducible \ads-module,
the corresponding on-shell field therefore qualifying for being an elementary field.
\begin{enumerate}
  \item[(i)] $\boldsymbol{E_0=d-q,\ p\geqslant q}\,$: There is a gauge symmetry with the gauge
   parameter having the symmetry type $[p,q-1]\,$. The corresponding gauge transformations read
   \begin{equation}
   \delta_{\epsilon} \phi^{a[p],b[q]} \;=\; D^{b}\epsilon^{a[p],b[q-1]}
                      + \frac{p\,(-1)^{p+1}}{(p-q+1)}\;D^{a}\epsilon^{a[p-1]b,b[q-1]}\quad ,
   \end{equation}
   where the last term is needed for the whole expression to have the $\mathfrak{gl}_d$ symmetry
   type $[p,q]\,$. The gauge parameter obeys equations analogous to
   (\ref{OnMassShellA})-(\ref{OnMassShellC}), but with a different value for the
   critical mass.
   The quotient Verma module is unitary. The field is called \emph{massless} \cite{Metsaev:1995re} because of the gauge
   transformations with one derivative, as occuring for a massless field in flat spacetime.
   It is that kind of field, with $q=1\,$, that we will be considering in the present paper.
  \item[(ii)] $\boldsymbol{E_0=d-p-1,\ p>q}\,$: That critical point corresponds to the second
   possibility for a gauge parameter, namely the one having the type $[p-1,q]\,$,
   \begin{equation}
      \delta_{\epsilon} \phi^{a[p],b[q]} \;=\; D^{a}\epsilon^{a[p-1],b[q]}\quad .
   \end{equation}
   The projector onto the symmetry type $[p,q]$ trivializes down to a single term.
   In this case, however, the quotient Verma module is non-unitary, which sorts out such
   massless fields in any reasonable field theory in \AdS.
   These types of fields are also called massless \cite{Metsaev:1995re} due to first order gauge transformation law.
   Due to the non-unitarity property, we will not be considering them in the present paper.
  \item[(iii)] $\boldsymbol{E_0=d-q-1,\ p=q}\,$: If one allows for higher derivatives in gauge
   transformations, then there can be a third possibility \cite{Skvortsov:2009zu} with the gauge parameter having the
   symmetry type $[p-1,p-1]\,$:
   \begin{equation}
    \delta_{\epsilon} \phi^{a[p],b[p]} \;=\; D^{a}D^b\epsilon^{a[p-1],b[p-1]} +
             \lambda^2 \,\eta^{ab}\,\epsilon^{a[p-1],b[p-1]}\quad ,
    \end{equation}
   where the second term above represents a lower-derivative $\Lambda$-correction to the first one.\footnote{In the case of
   a $[p,p,\ldots,p]$-type field, one has a tail of
   lower-derivative terms with increasing powers of $\lambda^2\,$. In the simplest $[p,p]$-type
   case where the leading terms in the gauge transformation bring in two covariant derivatives,
   one has only ${\cal O}(\lambda^2)$ corrections.}
   In $\AdS$ these fields give rise to non-unitary modules so that
   they will not be considered in this paper.
   They are called ``partially-massless'' \cite{Deser:2001us} because the gauge parameter
   has fewer indices in comparison with the massless case, and thus one may think of a
   gauge symmetry weakening.
   In general, partially-massless fields
   constitute a discrete chain interpolating between non-gauge massive fields with the highest
   number of physical degrees of freedom, and massless ones with the lowest number thereof.
   For more details
   and the description of partially-massless fields in the frame-like (or Cartan) formalism,
   we refer to \cite{Skvortsov:2006at}.
\end{enumerate}

In the massless $\Lambda\rightarrow0$ limit, all the critical values
for $E_0$ simultaneously go to zero and one has a degeneracy of
gauge symmetry types: the partially-massless cases do not exist any
more and one observes that the equations
(\ref{OnMassShellA})-(\ref{OnMassShellC}) become simultaneously
invariant under both types of gauge transformations with parameters
of type $[p,q-1]$ and $[p-1,q]\,$. Therefore, due to this gauge
symmetry enhancement, the Minkowski spin-$[p,q]$ field has fewer
degrees of freedom in comparison with any of the two massless fields
in anti-de Sitter background. See \cite{Brink:2000ag} and
\cite{Boulanger:2008up,Boulanger:2008kw} for extended discussions.
\vspace*{.2cm}

So far the description was on-shell. In order to give an off-shell description, the constraints
(\ref{OnMassShellB})-(\ref{OnMassShellC}) have to be relaxed both for the field and its gauge
parameter. The field content has to be enlarged with new fields associated
with certain nonvanishing traces of the $\mathfrak{gl}_d$ tensor $\phi^{\mu[p],\nu[q]}\,$.
In the frame-like, or Cartan, approach to gauge fields in $\AdS$ background,
powerful tools exist in order to tackle the interaction problem,
so we now review the Cartan approach to gauge fields in $\AdS$.

\paragraph{Generalized connections.} It has been known since \cite{MacDowell:1977jt, Stelle:1979aj}
that anti-de Sitter gravity can be understood as a Yang--Mills-like theory with the gauge algebra
being \ads
\begin{align}
\parbox{0.5cm}{\boldpic{\YoungAA}},\quad W^{A,B}_\mu dx^\mu &&\longrightarrow
&&\delta_\epsilon\phi_{\mu\nu}=D_\mu\epsilon_{\nu}+D_\nu\epsilon_{\mu}
\end{align}
where the gauge transformations above have been written for the linearized theory.
\vspace*{.3cm}

Vasiliev showed in \cite{Vasiliev:2001wa} that a free spin-$s$ gauge field can be described by
a one-form that, as an irreducible tensor of \ads{}, has the symmetry of a rectangular
two-row Young diagram of length-$(s-1)$:
\begin{align}\label{IntroDemotionSequence}
\parbox{2.3cm}{\boldpic{\RectBRow{6}{6}{$s-1$}{$s-1$}}},\quad
W^{A(s-1),B(s-1)}_\mu dx^\mu &&\longrightarrow &&
\delta_\epsilon\phi_{\mu(s)}=D_\mu \epsilon_{\mu(s-1)}\quad .
\end{align}
This lead
\cite{Alkalaev:2003hc, Alkalaev:2003qv, Skvortsov:2006at, Boulanger:2008up, Boulanger:2008kw}
to the study of the generalized connections (or Yang--Mills-like fields) of the anti-de Sitter
algebra that are defined by differential forms of arbitrary degrees taking their values in
arbitrary irreducible representation of \ads{}. In the most general cases, the relation between
any given gauge theory (not necessarily unitary) and the corresponding generalized
\ads{}-connections was given in \cite{Skvortsov:2009zu}, with the details left in
\cite{Skvortsov:2009nv}.

\paragraph{$\boldsymbol{AdS}$ background for generalized connections.}
Pick a generalized \ads{}-connection and denote it by  $W^\Ds\fm{q}$.
It is defined by the form degree $q$ and the finite-dimensional tensor module $\Ds$ of \ads.
The anti-de Sitter background can be described by a connection $\Omega^{A,B}$ that
is a one-form taking its values in the adjoint representation of \ads.
In order for it to describe \AdS{}, it must be a nondegenerate solution of
\begin{align}\label{FlatAdSConnection}
d\Omega^{A,B} + \Omega\fud{A,}{C}\wedge\Omega^{C,B} &=0 &&
\longleftrightarrow& \DO^2&=0\quad, \qquad \DO=d+\Omega\quad.
\end{align}
The linearized gauge theory with connection $W^\Ds\fm{q}$ is defined off-shell
by specifying gauge transformations and a gauge invariant curvature
\begin{align}
\delta_\epsilon W^\Ds\fm{q} & = \DO \epsilon^{\Ds}\fm{q-1}\quad,
\label{GenConnA}\\
R^\Ds\fm{q+1} & = \DO W^\Ds\fm{q}\quad, && \delta_\epsilon
R^\Ds\fm{q+1} = 0\quad, && \DO R^\Ds\fm{q+1} = 0\quad,
\label{GenConnB}
\end{align}
where both $\epsilon^{\Ds}\fm{q-1}$ and $R^\Ds\fm{q+1}$ take their values in the
same \ads-module $\Ds\,$.
At some point one may need to break the manifest \ads-symmetry of the construction to
extract the background vielbein $h^a_\mu$ and spin-connection $\varpi^{a,b}_\mu$
out of $\Omega^{A,B}\,$.
This can be achieved via introducing \cite{Stelle:1979aj, Vasiliev:2001wa} a normalized
vector $V^A V_A=-1\,$, called compensator,
which carries no propagating degrees of freedom.
The stability algebra of $V^A$ is identified with the local Lorentz subalgebra
$\lorentz\subset\ads\,$.
The appropriate \ads-covariant definitions of the background vielbein $H^A$
and Lorentz connection $\Omega_L^{A,B}$ have the form \cite{Vasiliev:2001wa}
\begin{align}
\lambda \, H^A & = \DO V^A \quad, & \Omega_L^{A,B} = \;\Omega^{A,B}- \lambda(V^AH^B-H^AV^B)\quad ,
\label{AdSDefinitions}
\end{align}
where $H^A_\mu$ is required to have the maximal rank $d$ in order to give rise
to a nondegenerate vielbein, a necessary condition entering the definition of the $\AdS$ background.
As a consequence of the above definitions, one has the following natural constraints on
$H^A\,$, $V^A\,$:
\begin{align}
H^BV_B & = 0 \quad,   &   DV^A  & = 0 \quad ,  &   DH^A & = 0 \quad .
\end{align}
in terms of the Lorentz covariant derivative $D$ defined by
\begin{equation}
 D = d + \Omega_L  \quad .
\end{equation}
Let the $(d+1)$th value of an $\ads$ index be denoted by the symbol $\bullet\,$,
so that $A=(a,\bullet)\,$.
Taking $V^A = \delta^A_\bullet\,$, the definitions (\ref{AdSDefinitions}) yield
\begin{align}
\lambda\,H^a_\mu & = {\Omega_\mu}^{a,}_{\phantom{a,}\bullet} \quad, &
H^\bullet_\mu & = 0 \quad, &
\Omega_L^{a,b} & = \Omega^{a,b}\quad.
\end{align}
Therefore (\ref{AdSDefinitions}) is an $\ads$-covariant way to decompose an
antisymmetric matrix $\Omega^{A,B}$ into a vector to be identified with $h^a$
and an antisymmetric matrix of a lower rank to be identified with
$\varpi^{a,b}\,$. For a somewhat more geometric presentation of the above
material, see e.g. \cite{Boulanger:2008kw}.

\paragraph{Lorentz view on generalized connections.} Upon decomposing $\Ds$ of $W^\Ds\fm{q}$ into
irreducible Lorentz modules with the help of the embedding of \lorentz{} into \ads{} defined by
$V^A\,$, the $\ads$-connection $W^\Ds\fm{q}$ gives rise to a set of $\lorentz$-fields
which are certain generalizations of the dynamical vielbein and Lorentz connection.

To put the theory on-shell one may impose equations by setting certain components of
$R^\Ds\fm{q+1}$ to zero. Similarly to the case of gravity theories with vanishing torsion,
the generalized spin-connections are expressed in terms of derivatives of the vielbein field.
For a spin-$s$ field the appropriate on-mass-shell condition reads\cite{Vasiliev:2001wa}
\begin{align}
R\fm{2}^{A(s-1),B(s-1)} & = H_M H_N \, C\fm{0}^{A(s-1)M,B(s-1)N}\quad, &
C\fm{0}^{A(s),B(s-1)C}V_C & =0\quad ,
\label{SpinsOnmassshell}
\end{align}
where the zero-form $C\fm{0}^{A(s),B(s)}$ is called the spin-$s$ Weyl tensor.
It is an irreducible tensor of \ads{} type $(s,s)\,$.
The condition that the Weyl tensor be $V^A$-transversal means that it is effectively
an irreducible \lorentz{}-tensor of the same type $(s,s)$.
The curvature two-form $R\fm{2}^{A(s-1),B(s-1)}$ decomposes under $\lorentz\subset\ads$ into the
following set of two-forms
\begin{align}
R\fm{2}^{A(s-1),B(s-1)} \;\leftrightsquigarrow \;
\left\{R^{a(s-1)}\fm{2}\;, \; R^{a(s-1),b}\fm{2}\;,\; \ldots\;,\;
R^{a(s-1),b(s-2)}\fm{2}\;, \;R^{a(s-1),b(s-1)}\fm{2} \right\} \quad .
\end{align}
Therefore, the equations (\ref{SpinsOnmassshell}) set all but one Lorentz components of
the curvature two-form to zero.
For a spin-two dynamical field described by $e^a$ and $\omega^{a,b}\,$,
this gives the linearized Einstein equations with a cosmological constant term:
\begin{align}
\DL e^a +\omega\fud{a}{c}\,h^c & =0 \quad,
\label{spintwoomsA}\\
\DL \omega^{a,b} + \lambda^2 \, h^a e^b + \lambda^2 \, e^a h^b & = h_m h_n \, C^{am,bn}\quad .
\label{spintwoomsB}
\end{align}
It should be noted that the spin-two and the generalized spin-$s$
Weyl tensors are differentially constrained, in the sense that they satisfy differential
Bianchi identities following from $\DO R^\Ds\fm{q+1}\equiv0\,$.

The problem of identifying the on-shell physical fields, the Weyl tensors and
equations of motion can be reduced to a cohomological problem
\cite{Lopatin:1987hz, Shaynkman:2000ts, Alkalaev:2003qv, Skvortsov:2006at, Boulanger:2008up,
Boulanger:2008kw} with the answer known in full generality \cite{Skvortsov:2009nv}.

\paragraph{Generalized connections for $[k,q]$-type gauge fields.} As was shown
in \cite{Skvortsov:2009zu} in full generality --- the unitary cases
were previously discussed in \cite{Alkalaev:2003hc}--- a family of
spin-$[p,r]$ fields in $\AdS$ is described by generalized gauge
connections of the form
\begin{equation}
W^{A[k+1]}\fm{q} \equiv W^{A_1...A_{k+1}}_{\mu_1...\mu_q}\;dx^{\mu_1}\wedge...\wedge dx^{\mu_q},
\qquad \mbox{antisymmetric in\ } A_1...A_{k+1}\quad,
\end{equation}
where the two parameters $k$ and $q$ run over all admissible values
($q\geqslant 1$ since we are working with gauge fields).
The precise correspondence \cite{Skvortsov:2009zu} is given below.
\vspace*{.3cm}

In Table \ref{kqtable} we summarize the main features of the theories for the
various $[p,r]$-type gauge fields. From the discussion at the beginning of the
present section it is obvious that to specify a gauge theory in \AdS{} it is
enough to specify the spin of the field and the symmetry type of its gauge
parameter, or equivalently the spin of the field and the symmetry of its primary
Weyl tensor.
%
\begin{table}[h]
\begin{tabular}{|c||c|x{3.5cm}|c|c|c|}\hline
$k$ and $q$ & unitarity & massless(m.)/ p.-massless(p.m.) & field & gauge parameter
& Weyl tensor
\tabularnewline\hline\hline
$\boldsymbol{k+1=0}$  & yes   & m.  & \OneColumn{2}{q} &  \OneColumn{1}{q\!-\!1}
& \OneColumn{3}{q\!+\!1}
\tabularnewline\hline
$\boldsymbol{k>q}$& yes   & m.  & \TwoColumn{4}{2}{k}{q}&  \TwoColumn{3}{1}{k}{q\!-\!1}
& \TwoColumn{4}{3}{k}{q\!+\!1}
\tabularnewline\hline
$\boldsymbol{k=q}$& yes   & m.    & \TwoColumn{2}{2}{q}{q}&  \TwoColumn{2}{1}{q}{q\!-\!1}
& \TwoColumn{3}{3}{q\!+\!1}{q\!+\!1}
\tabularnewline\hline
{$\boldsymbol{k=q-1}$}  & no  & p.m.  & \TwoColumn{2}{2}{q}{q} &  \TwoColumn{1}{1}{q\!-\!1}{q\!-\!1}
& \TwoColumn{3}{2}{q\!+\!1}{q}
\tabularnewline\hline
{$\boldsymbol{k<q-1}$}  & no    & m.    & \TwoColumn{3}{1}{q}{k} &  \TwoColumn{2}{1}{q\!-\!1}{k}
& \TwoColumn{4}{1}{q\!+\!1}{k}
\tabularnewline\hline
\end{tabular}
\caption{Fields, gauge parameters and primary Weyl tensors for the various possible
$[k,q]$-type gauge fields in $\AdS\,$.}
\label{kqtable}
\end{table}
%
The various case depicted in Table \ref{kqtable} are described as follows:
\begin{description}
  \item[$\boldsymbol{k+1=0: }$] Gauge $q$-forms, i.e. massless fields with spin
    given by $[q]\,$.
  \item[$\boldsymbol{k=q: }$] Unitary massless fields of spin $[q,q]$.
    These fields are very close to the Minkowski massless fields. The graviton,
    whose spin is
    $[1,1]\,$, belongs to this class.
  \item[$\boldsymbol{k>q: }$] Unitary massless fields with spin-$[k,q]\,$.
    The gauge parameter has spin $[k,q-1]\,$.
  \item[{$\boldsymbol{k=q-1: }$}] Nonunitary partially massless field of spin
   $[q,q]$ and depth two, i.e.
    the gauge parameter has spin $[q-1,q-1]$ and the gauge transformations
    contain two derivatives.
  \item[{$\boldsymbol{k<q-1: }$}] Nonunitary massless fields of spin $[q,k]\,$.
    The gauge parameter has the symmetry type $[q-1,k]\,$.
\end{description}

\paragraph{Lorentz view on two-column fields.} We restrict ourselves to the unitary cases only,
investigated in great details in \cite{Alkalaev:2003hc}.
The generalized gauge connection we will consider
\begin{align}
W^{A[k+1]}\fm{q}\quad,
\end{align}
decomposes into two Lorentz connections
\begin{align}
e^{a[k]}\fm{q}&=W^{a[k]\bullet}\fm{q} &&\mbox{and}&& \omega^{a[k+1]}\fm{q}=W^{a[k+1]}\fm{q}
\quad.
\end{align}
After some $\lambda$-rescalings, the curvature $R^{A[k+1]}\fm{q+1}$
in Lorentz components reads
\begin{align}
R^{a[k]}\fm{q+1} &\;=\; \DL e^{a[k]}\fm{q} - h_m\,\omega^{ma[k]}\fm{q}\quad,
\label{CurvatureA}\\
R^{a[k+1]}\fm{q+1} &\;=\; \DL \omega^{a[k+1]}\fm{q}-\lambda^2 h^a e^{a[k]}\fm{q}\quad,
\label{CurvatureB}
\end{align}
and is invariant under gauge transformations of the form\footnote{To see this one has
to use $\DL^2 \xi^a=-\lambda^2 h^a \wedge h_m\, \xi^m$ for any vector $\xi^a$.}
\begin{align}
\delta_\epsilon e^{a[k]}\fm{q}&\;=\;\DL \epsilon^{a[k]}\fm{q-1} - h_m\,\epsilon^{ma[k]}\fm{q-1}
\quad,
\label{GaugeTrA}\\
\delta_\epsilon \omega^{a[k+1]}\fm{q}&\;=\;\DL \epsilon^{a[k+1]}\fm{q-1}-\lambda^2 h^u
\epsilon^{a[k]}\fm{q-1}
\quad.
\end{align}
The potential $\varphi^{a[k],b[q]}$ is to be identified with the corresponding component of the
generalized vielbein $e^{a[k]}\fm{q}\,$. This can be easily done with a choice of the symmetric
notation for the field
\begin{align}
&&\varphi^{a[k],b[q]}&&\longrightarrow&&
\varphi^{\overbrace{\scriptstyle aa,bb,...,cc}^{\scriptstyle q\mbox{\scriptsize\ pairs}},
\overbrace{\scriptstyle u,v,...,w}^{\scriptstyle k-q}}&&
\end{align}
then the potential is a maximally symmetric part of the vielbein
\begin{align}
\varphi^{aa,bb,...,cc,u,...,w}&=e^{ab...cu...w|ab...c}, & e^{a[k]|u...w}&=e^{a[k]}_{\mu_1...\mu_q}
h^{\mu_1 u}...h^{\mu_q w},
\end{align}
where $h^{\mu c}$ is the inverse of the background vielbein $h_\mu^a\,$:
$\,h^{\mu c} h_\mu^a=\eta^{ac}\,$. One has to use $h^{\mu c}$ or $h_\mu^a$ in order to interpret
the gauge connections in terms of the potentials.
The fiber version of (\ref{GaugeTrA}) reads
\begin{align}
\delta_\epsilon e^{a[k]|v[q]} &= \DL^v \epsilon^{a[k]|v[q-1]} - \epsilon^{va[k]|v[q-1]}.
\end{align}
The second term is a shift (St\"uckelberg-like) symmetry, a kind of local Lorentz transformations
for the vielbein, whose purpose is to remove the unwanted components of the vielbein in order for it
to match with the content of the field potential.
Indeed, one observes that the second term does not shift $\varphi\,$.
The same kind of shift symmetry acts on the gauge parameters too, as a result of reducibility of
gauge transformations:
\begin{align}
\delta_\epsilon W^{\Ds}\fm{q}&\equiv0 && \mbox{for} && \epsilon^\Ds\fm{q-1}=\DO\chi^\Ds\fm{q-2}
\quad.
\end{align}
This is exactly the cohomological origin\footnote{This is so-called \Sigm-
cohomology \cite{Lopatin:1987hz}, which have been one of the main technical tools
in many works, e.g. \cite{Shaynkman:2000ts,Vasiliev:2007yc,Boulanger:2008kw,Skvortsov:2009nv}.}
of the problem of finding physically relevant components
in $\epsilon^\Ds\fm{q-1}$, $W^\Ds\fm{q}$ and $R^\Ds\fm{q+1}\,$.
Let us denote by $\Sigm$ the operator taking some Lorentz $[k+1]$-type $q$-form
$B^{a[k+1]}\fm{q}$ to the Lorentz $[k]$-type $(q+1)$-form $B^{a[k]}\fm{q+1}=h_m B^{ma[k]}\fm{q}\,$
\begin{align}
 & 0&&\longrightarrow&&B^{a[k+1]}\fm{q} && \xrightarrow{\quad\Sigm\quad}  && B^{a[k]}\fm{q+1} && \longrightarrow && 0
\quad.
\end{align}
Then the cohomology groups $H^{q-1}(\Sigm)$ and $H^{q}(\Sigm)$ correspond to the differential
(as opposed to  St\"uckelberg) gauge parameters and the potential $\varphi^{a[k],b[q]}\,$,
including all the traces that are necessary to formulate the theory off-shell.
By solving the cohomology problem, which is very simple in this case, one concludes that the
symmetries of the potential and gauge parameter match the required ones.
\vspace*{.3cm}

The problem of identifying the content of the curvature is more subtle and requires the use of
(\ref{CurvatureA})-(\ref{CurvatureB}) together with the Bianchi identities
$\DO R^\Ds\fm{q+1}\equiv0\,$:
\begin{align}
\DL R^{a[k]}\fm{q+1} - h_m\, R^{m a[k]}\fm{q+1} & \equiv 0 \quad,
\label{tallBianchiA}\\
\DL R^{a[k+1]}\fm{q+1} + \lambda^2 \,h^a\, R^{a[k]}\fm{q+1} &\equiv0\quad .
\label{tallBianchiB}
\end{align}
Actually, $H^{q+1}(\Sigm)$ gives all linearly independent, gauge-invariant by construction,
components of the curvatures.
The \Sigm-exact pieces correspond to those components that can be shift to zero by redefining
$\omega^{a[k+1]}\fm{q}\,$. The equations that do not belong to $H^{q+1}(\Sigm)$ either express
certain components of $\omega^{a[k+1]}\fm{q}$ in terms of derivatives of $\varphi\,$,
or are the differential consequences of more fundamental equations that lie in $H^{q+1}
(\Sigm)$\footnote{Evidently, taking a derivative of a gauge-invariant equation produces
one more gauge-invariant equation of higher order.}.

Which representatives of $H^{q+1}(\Sigm)$ should be set to zero depends on the dynamics one wishes
to describe.
For the example of gravity, if we set the Weyl tensor to zero keeping the Ricci tensor
unconstrained, then the equations will describe various conformally flat backgrounds
\cite{Vasiliev:2007yc} (in $d>3$ the manifold is locally conformally flat if and only if
the Weyl tensor vanishes). Instead, if one only keeps nonzero the traceless part
of the curvature tensor (\emph{i.e.} the Weyl tensor), then one
effectively imposes the Einstein equations.

Generally, for $W^\Ds\fm{q}$ to describe an irreducible propagating field one should set
all the components of the curvature to zero but the generalized Weyl tensor.
To give a concrete example, we will consider a spin-$\parbox{12pt}{\smallpic{\YoungBA}}$
field $\varphi\sim [2,1]\,$, \emph{i.e.}
$k=2$, $q=1\,$, represented in the one-form $W^{A[k+1]}\fm{1}\,$,
but otherwise all the statements hold true for any spin-$[k,q]$ field.

The space of gauge-invariant candidate equations, i.e. the representatives of
$H^{q+1}(\Sigm)$ for $\Ds$ having type $[k+1]\,$, is parameterized as follows
\cite{Boulanger:2008kw,Skvortsov:2009nv}.
The (primary) Weyl tensor is to be extracted from the components of
$R^{a[k]}\fm{q+1}\,$.
It is a traceless tensor of type $[k,q+1]\,$ (see also the last column of
Table \ref{kqtable}):
\begin{align}
\parbox{20pt}{\YoungBB} && \longleftrightarrow && \mbox{traceless part of\ }
\left(2\,D^{(a} e^{a)(b|b)} + 2\,D^{(b} e^{b)(a|a)} = D^a\varphi^{bb,a}+D^b\varphi^{aa,b}\right)
\quad.
\end{align}
Indeed, one can see that the \Sigm-exact part $\Sigm(\omega)$ of the curvature in
(\ref{CurvatureA}) drops out for such a permutation of indices.
The above tensor, of first order in derivatives of the potential field,
is a representative of $H^{q+1}(\Sigm)\,$.
There are more representatives of $H^{q+1}(\Sigm)$ in $R^{a[k]}\fm{q+1}$ that are
collectively given by a traceful tensor with the symmetry $[k-1,q]$ called the
torsion $T^{a[k-1],b[q]}\,$:
\begin{align}
\parbox{22pt}{\YoungB}\oplus\bullet && \longleftrightarrow &&
  D^{(a}_{\phantom{a}} e\fud{a)m|}{m} + D_m e^{m(a|a)} = D^{(a}_{\phantom{a}}\varphi\fud{a)m,}{m}-D_m\varphi^{aa,m}
\quad.
\end{align}
The reason to call it \emph{torsion} is that it is a part of the
($\DL e+\omega$)-like curvature that coincides with the torsion for spin-two.
There is no nontrivial representative of the $\Sigm$-cohomology in the spin-two
torsion tensor, of course, as $T_{\mu\nu}{}^a=0$ does not restrict the spin-two
vielbeins $e^a\,$ but instead expresses the spin connection in terms of them.

The intersection of $H^{q+1}(\Sigm)$ with $R^{a[k+1]}\fm{q+1}$ is empty.
However, as we keep the primary Weyl tensor nonzero, we have to keep all its
descendants (secondary Weyl tensor, \emph{etc.}) that are produced by taking
various appropriately projected derivatives of it.
The first such descendant, the secondary Weyl tensor, is a traceless
type-$[k+1,q+1]$ component of $R^{a[k+1]}\fm{q+1}$. All other types of tensors
appearing as first derivative of the primary Weyl tensor or the torsion must be
set to zero by imposing equations of motion. These are given by a tensor
$G^{a[k],b[q]}$ that has the same $\mathfrak{gl}_d\,$-symmetry and trace
properties
as $\varphi^{a[k],b[q]}\,$.
As a differential expression, $G^{a[k],b[q]}[\varphi]$ defines a second-order
operator starting like $\square \varphi^{a[k],b[q]} +\ldots\,$.
It represents the equations of motion.
Actually, one apparently still needs to impose $T=0$ to get an irreducible
module\footnote{The constraint $T=0$ is the price one has to pay in $AdS$
because one of the two gauge symmetries of massless spin-$[k,q]$ field in
Minkowski space gets broken in $\AdS\,$.}.
Fortunately, for a two-column spin-$[k,q]$ field the problem is cured by noting
that $D_m G^{a[k-1]m, b[q]}\equiv\Lambda T^{a[k-1],b[q]}\,$.
Hence, once the second-order differential equations $G[\varphi]=0$ are imposed,
$T=0$ becomes a consequence of it\footnote{This is similar to $\pl^\mu A_\mu=0$
that appears as a consequence of the Proca field equation
$\pl^\mu(\square A_\mu-\pl_\mu\pl^\nu A_\nu+m^2A_\mu)=0\,$.
Let us note that the case of fields with type $[k,q,q,...,q],$,
$k\geqslant q\,$, is degenerate. In general, one cannot achieve $T=0$ by acting on $G=0$ with
derivatives, so that $T=0$ is an independent equation that must be imposed.}.

To summarize, the necessary equations of motion that leave the primary Weyl tensor
and its descendant free read
\cite{Alkalaev:2003hc,Boulanger:2008up,Boulanger:2008kw,Skvortsov:2009zu}
\begin{align}
R^{a[k]}\fm{q+1}&=h_m...h_m\, C^{a[k],m[q+1]}\,, &
R^{a[k+1]}\fm{q+1} & = h_m...h_m \,C^{a[k+1],m[q+1]}\label{tallomsA}
\end{align}
where the zero-form $C^{a[k],m[q+1]}$ is the primary Weyl tensor, while its descendant,
the secondary Weyl tensor $C^{u[k+1],m[q+1]}$, is given by a derivative of the primary Weyl tensor
as can be seen from (\ref{tallBianchiA})-(\ref{tallBianchiB}).
Both zero forms are irreducible Lorentz tensors.
Equivalently, one can use manifestly \ads-covariant form of equations
\begin{align}
R^{U[k+1]}\fm{q+1}&=H_M...H_M \,C^{U[k+1],M[q+1]}\,, &
C^{U[k+1],M[q]B}V_B&\equiv0\label{tallomsB}
\end{align}
where the \ads-irreducible zero-form $C^{U[k+1],M[q+1]}\,$, subject to {\textit{non}}-complete
$V$-transversality condition, contains two Lorentz components corresponding to the primary and
secondary Weyl tensors $C^{a[k],m[q+1]}$ and $C^{a[k+1],m[q+1]}\,$.

\paragraph{Action.} The next problem is to find an action yielding $G^{a[k],b[q]}[\varphi]=0\,$.
We will see that the use of the generalized gauge connections together with the rules
of exterior differential algebra leave almost no freedom here
as compared to the metric-like tensor $\varphi^{a[k],b[q]}$
for which one can write many terms with two derivatives as an Ansatz for the Lagrangian.
\vspace*{.3cm}

\noindent Let us define the following two volume forms in \AdS- and Lorentz-covariant formulations:
\begin{align*}
\mbox{vol}_{U[k]}&=\epsilon_{U_1...U_k B_{k+1}...B_{d}W}\,
H^{B_{k+1}}\wedge ...\wedge H^{B_{d}} V^W\,, &
\mbox{vol}_{u[k]}&=\epsilon_{u_1...u_k b_{k+1}...b_{d}}\,
h^{b_{k+1}}\wedge ...\wedge h^{b_{d}}\quad .
\end{align*}
The volume form $\mbox{vol}_{u[k]}$ obeys the identity
\begin{align}
\label{MSIdentityVielbeins}
h^c\wedge \mbox{vol}_{u_1...u_k } \;=\;
\frac{(-1)^k}{(d-k+1)}\;\sum_{i=1}^{i=k}(-1)^{i}\delta^c_{u_i}\,\mbox{vol}_{u_1...\hat{u}_i...u_k}
\quad,
\end{align}
and a similar identity for $\mbox{vol}_{U[k]}\,$.

The most general parity-even, manifestly gauge-invariant Ansatz for the action
reads
\cite{Alkalaev:2003hc}
\begin{align}
S&=\frac1{\lambda^2}\int \left(
a_1 R^{u[q+1]a[k-q-1]}\wedge R\fud{u[q+1]}{a[k-q-1]} +
a_2R^{u[q+1]a[k-q]}\wedge R\fud{u[q+1]}{a[k-q]}\right)
\wedge \mbox{vol}_{u[2q+2]}\nonumber\\
&\qquad\qquad+\frac\alpha{\lambda^2}\int \DL\left(R^{u[q]a[k-q]}\wedge
R\fud{u[q+1]}{a[k-q]}
\wedge \mbox{vol}_{u[2q+1]}\right)\quad,
\label{HookActionA}
\end{align}
where the second line includes a boundary term.
In manifestly \AdS-covariant terms:
\begin{align}
S & = \frac1{\lambda^2}\;\int \left(
\alpha_1 R^{U[q+1]A[k-q-1]C}\wedge R\fudu{U[q+1]}{A[k-q-1]}{C}V_CV_C +
\alpha_2R^{U[q+1]A[k-q]}\wedge R\fud{U[q+1]}{A[k-q]}\right)\nonumber\\
&\qquad\qquad\wedge\mbox{vol}_{U[2q+2]} +
\frac\alpha{\lambda^2}\int \DO\left(R^{U[q]A[k-q]C}V_C\wedge
R\fud{U[q+1]}{A[k-q]}
\wedge\mbox{vol}_{U[2q+1]}\right)\quad.
\label{HookActionB}
\end{align}

It consists of two terms and one boundary term
\cite{Alkalaev:2003hc} that can be used to adjust the ratio
$\frac{\alpha_1}{\alpha_2}\,$ at will. As we will see, in order to
switch on the gravitational interactions, the ratio
$\frac{\alpha_1}{\alpha_2}\,$ has to be fixed in the right way.

It appears that $d\geqslant 2q+2$ must be true in order for the action to make sense.
For $d$ odd it automatically follows that there is no action for a spin-$[k,k]$ field with
$k=\frac{d-1}2\,$. The rectangular spin diagrams with height equal to the rank of \msv,
the Wigner little algebra in \AdS, correspond to the higher-spin singletons
(or doubletons for $d=5$),
see e.g. \cite{Iazeolla:2008ix,Bekaert:2009fg} for some recent works.
A singleton is an irreducible representation of \ads{} that is too
short to possess any bulk degrees of freedom in \AdS. It is not possible to write the bulk
action for singletons, which is manifested in $d\geqslant 2q+2\,$.

The Lagrangian equations of motion
\begin{align}
R^{u[q]a[k-q]}\wedge
\delta\omega\fud{u[q+1]}{a[k-q]}\wedge\mbox{vol}_{u[2q+1]}&=0\\
R^{u[q+1]a[k-q]}\wedge\delta e\fud{u[q]}{a[k-q]}\wedge\mbox{vol}_{u[2q+1]}&=0
\end{align}
express $\omega$ as $\DL\varphi$ and then impose $G^{a[k],b[q]}[\varphi]=0$, as
required.

%
\section{Fradkin-Vasiliev cubic interactions}\setcounter{equation}{0}
\label{sec:FV}

In this section we first review the Fradkin--Vasiliev procedure
\cite{Fradkin:1986qy} (see also \cite{Vasiliev:2001wa,
Alkalaev:2002rq, Alkalaev:2010af}), putting emphasis on the
situation where a higher-spin algebra $\mathfrak{g}$ containing
$\ads$ and decomposable under $\ads$ is given from the outset.

Given a quadratic (free) action $S_0=S_0[\varphi]$ invariant under abelian gauge
transformations $\delta\varphi=\delta_0\varphi\,$, the general perturbative procedure
for finding cubic and higher interaction vertices leads to considering a formal expansion
in powers of some coupling constant $g$
\begin{align}
S & = S_0 + g \,S_1 + {\cal O}(g^2)\quad, &
\delta\varphi& =\delta_0\varphi + g\, \delta_1\varphi
+ {\cal O}(g^2) \quad,
\end{align}
and look order by order in powers of $g$ for solutions of
\begin{align}
0 = \delta S = \delta_0 S_0 + g\,\left(\delta_0 S_1+\delta_1 S_0\right) +{\cal O}(g^2)\quad,
\end{align}
where the order-$g^0$ terms vanish by gauge invariance of the quadratic action $S_0\,$.
The cubic interaction vertices are governed by terms of order $g\,$,
\begin{align}
0 \;=\; \delta_0 S_1 + \delta_1 S_0 \;=\; \delta_1\varphi\,\frac{\delta S_0}{\delta\varphi}
+\delta_0\varphi\,\frac{\delta S_1}{\delta\varphi}\quad.
\label{FVcond}
\end{align}
Taking into account that $\frac{\delta S_0}{\delta\varphi}$ are the linear equations of motion,
the problem of cubic interactions is reduced to finding $S_1$ such that its gauge variation
under abelian transformation $\delta_0\varphi$ vanishes on-mass-shell,
\begin{align}
\left.\frac{\delta S_1}{\delta\varphi}
\delta_0\varphi\right|_{\frac{\delta S_0}{\delta\varphi}=0}=0\quad,
\label{FVcondA}
\end{align}
\emph{i.e.} is proportional to $\frac{\delta S_0}{\delta\varphi}\,$.
Having achieved this, $\delta_1\varphi$ can be extracted
by inspection of the terms $\delta_1\varphi\,\frac{\delta S_0}{\delta\varphi}$
in (\ref{FVcond}). Now this general consideration will be specified to higher-spin theories following the pioneering work \cite{Fradkin:1986qy}.
\vspace*{.3cm}

Suppose there is a set of generalized one-form connections
$\omega=\left\{ W^k\fm{1}\right\}\,$,
$k=1,...\,$, collectively denoted by $\omega$, with the $\ads$-connection
$W^{A,B}\fm{1}$ belonging to the set.
Suppose also that there is an associative algebra structure
$\mathfrak{g}$ (with product denoted by $\star$ in the sequel) on the set
$\{W^k\}$ such that $\ads\subset\mathfrak{g}$ acts on $\mathfrak{g}$ via the
adjoint action.
Then $\mathfrak{g}$ may be called a higher-spin algebra
\cite{Konshtein:1988yg, Vasiliev:2004cm}.
There is a well-defined linear gauge theory (\ref{GenConnA})-(\ref{GenConnB})
on individual components $W^k\fm{1}$ of $\omega$, i.e. the spectrum of
$W^k\fm{1}$'s is given by decomposing $\omega$ of $\mathfrak{g}$ into
$\ads$-submodules $W^k\,$.
This linear gauge theory can be understood as a linearization of
$R=d\omega+\omega\star\omega\,$,
$\delta \omega=d\epsilon+[\omega,\epsilon]_\star$
over the anti-de Sitter background defined by $\Omega^{A,B}\,$,
(\ref{FlatAdSConnection}):
\begin{align}
R &= R_0+gR_1 = D_0\omega + g\,\omega\star\omega
\quad,\label{FullCurvature}\\
\delta\omega &= \delta_0 \omega + g\,\delta_1 \omega
= D_0\epsilon  + g\,[\omega,\epsilon]_\star
\quad, \\
\delta R &= \delta_0 R_0 + g\, (\delta_1 R_0 + \delta_1 R_0) + g^2\,\delta_1 R_1
 = 0 + g\,[R_0,\epsilon]_\star
+g^2\,[R_1,\epsilon]_\star
\quad.
\label{FVgaugeAC}
\end{align}
In terms of individual components one has
\begin{align}
R^k &= D_0 W^k + g \sum_{m,n} W^n \star W^m\quad,\\
\delta W^k &= D_0\epsilon^k + g \sum_{m,n} [W^m, \epsilon^n]_\star
\quad,\\
\delta R^k &=
g \sum_{m,n} [ {{R^m_0}},\epsilon^n]_\star
+ {\cal{O}}(g^2)\quad.
\label{FVgaugeC}
\end{align}
One then tries to find a quadratic action $S_0$ for individual fields
$W^k\in\omega\,$, bilinear in the linearized curvatures $R_0^k$
($H$ is the background tetrad $H^A$):
\begin{align}
S_0&=\sum_k \alpha_k \,S_0^k\,,& S_0^k{[\omega,V]} &= \frac12 \int_{M_d} \left(R^k_0 \wedge
R^k_0\right){}^{\!\!ABCD} \wedge \mbox{vol}_{ABCD}\,,
\label{FVaction}
\end{align}
where $\left(R^k_0 \wedge R^k_0\right)^{[ABCD]}$ means that all but four indices
carried by curvatures are contracted among themselves or with a number of
compensators $V^A\,$. In general there may be a lot of such nontrivial
contractions that can contribute, e.g. two in (\ref{HookActionB}).
It is the problem of constructing quadratic action to determine all free
coefficients inside $S^k_0$ up to an overall factor $\alpha_k\,$.
The indices $ABCD$ appear implicitly antisymmetrized because they are
contracted with $\mbox{vol}_{ABCD}\,$.
The integrand is a $d$-form as is required, this is why $\mbox{vol}_{ABCD}$
has to carry four free indices.
Note that according to Table 1 none of the nonunitary fields can be given a Lagrangian as they are described
by gauge connections that do not have enough antisymmetric indices as compared to the form degree. Hence the FV procedure cannot be applied directly to nonunitary type-[p,q] fields.

The action $S^k_0$ is manifestly gauge invariant.
It is a generalization of the Stelle--West action \cite{Stelle:1979aj}.
The coefficients $\alpha_k$ account for possible different choices of
normalization for each individual quadratic action entering the sum.
The action $S^k_0$ for a spin-$s$ field was found in \cite{Vasiliev:2001wa}.
For two-column gauge fields, quadratic actions were elaborated along those
lines in \cite{Alkalaev:2003hc,Alkalaev:2005kw}.

The idea of Fradkin and Vasiliev \cite{Fradkin:1986qy}
was to use the same Ansatz for a cubic action, \emph{i.e.} to replace $R_0$
with the full Yang--Mills-like curvature $R$ and consider the
action modulo terms of order higher than cubic\footnote{It might seem that background tetrads $H^A$ that are hidden in $\mbox{vol}_{ABCD}$ has to be replaced with dynamical ones.
Fortunately \cite{Fradkin:1986qy}, the gauge variation of such terms can always be compensated. Therefore, it is everywhere implied that $\mbox{vol}_{ABCD}$ is a volume element with respect to the background. }
\begin{align}
S{[\omega,V]} = \frac12\,\sum_k \alpha_k \int_{M_d} \left(
R^k \wedge R^k\right){}^{\!\!ABCD} \wedge \mbox{vol}_{ABCD}
+\mathcal{O}(\omega^4)\quad.
\label{FVactionB}
\end{align}
Firstly, in order to have a nonabelian gauge algebra,
we make appear in the transformation laws
$\delta_1\varphi$ a part denoted $\delta_1^{hs}\varphi$ associated with
the algebra $\mathfrak{g}\,$,
plus some extra contribution $\delta_1^{ex}\,$ such that
$\delta_1\varphi=\delta_1^{hs}\varphi+\delta_1^{ex}\varphi\,$.
As a result (\ref{FVcondA}) reads
\begin{align}
\left.\left(\delta^{hs}_1\varphi\,\frac{\delta S_0}{\delta\varphi}
+\delta_0\varphi\,\frac{\delta S_1}{\delta\varphi}\right)
\right|_{\frac{\delta S_0}{\delta\varphi}=0}=0\label{FVcondB}
\end{align}
and it is $\delta_1^{ex}\varphi$ that has to be extracted once the solution to
(\ref{FVcondB}) is found. The most complicated work is to extract
$\delta_1^{ex}\varphi\,$.
Fortunately, whatever $\delta_1^{ex}\varphi$ is, the vertex is constructed once
(\ref{FVcondB}) is solved for $S_1\,$, so in practice one does not need to
struggle with finding $\delta_1^{ex}\varphi\,$ if one is only interested in
the vertex and not in the complete expression for the gauge transformations.
For example, in the case of pure gravity the diffeomorphism
$\delta_\xi$ along the vector field $\xi=\xi^\mu\partial_\mu$ is equivalent to a combination of
gauge transformations with
$\epsilon^a=\xi^\nu e^a_\nu$, $\epsilon^{a,b}=\xi^\nu \omega^{a,b}_\nu$
and a curvature-dependent term, e.g.
\begin{align}
\delta_\xi e^a_\mu &= \delta_{\epsilon}e^a_\mu - \xi^\nu
(d e^a +\omega\fud{a,}{b}\wedge e^b)_{\mu\nu}
&\delta_\epsilon e^a& =
d\epsilon^a-\epsilon\fud{a,}{b} \,e^b+\omega\fud{a,}{b}\,\epsilon^b\,.
\end{align}
The first term of the first equation represents $\delta_1^{hs}$ while the second
one corresponds to $\delta_1^{ex}$ and does not have a nice form in general.
For higher-spin fields $\delta_1^{ex}$ is much more
complicated \cite{Fradkin:1986qy}.
\vspace*{.3cm}

Having Eq. (\ref{FVgaugeC}) in mind, the gauge variation of
(\ref{FVactionB}) under $\mathfrak{g}$ is easy to evaluate:
\begin{align}
{\delta^{\mathfrak{g}} S} = \delta_0 S_1+\delta^{hs}_1 S_0=
g\sum_k \alpha_k\sum_{m,n} \int \left( R^k_0 \wedge [{R_0^m},
\epsilon^n]_\star
\right){}^{\!\!ABCD} \wedge \mbox{vol}_{ABCD} + \mathcal{O}(\omega^3\epsilon)
\quad.
\label{FVactionC}
\end{align}
To find cubic interactions among elementary fields sitting in $\omega$ one needs
to adjust $\alpha_k$ such that (\ref{FVactionC}) vanishes on free shell, up to
terms $\mathcal{O}(\omega^3\epsilon)$.
A simplification results from the central
on-mass-shell theorem, originally formulated for
$4d$ higher-spin fields in \cite{Vasiliev:1988sa}.
It turns out that almost all components of $R_0$ are zero on free mass-shell
except for some of them parameterized by the primary Weyl tensors (and certain
of its descendants for mixed-symmetry fields), \emph{i.e.}
\begin{align}
\left.\rule{0pt}{16pt}R^k_0\right|_{\frac{\delta S_0}{\delta\varphi}=0} =
H\wedge...\wedge H \;C^k\,, \label{omsgeneral}
\end{align}
where the generalized Weyl tensors collectively denoted by $C$ carry some indices
that are contracted with a number of background vielbeins to match the symmetry
type and form degree on both sides. The explicit examples are presented in
(\ref{SpinsOnmassshell}), (\ref{spintwoomsA})-(\ref{spintwoomsB}),
(\ref{tallomsA}) and (\ref{tallomsB}).

Having replaced $R^k_0$ with Weyl tensors on account of the central
on-mass-shell theorem (\ref{omsgeneral}), one can use the identity
(\ref{MSIdentityVielbeins}) that basically gives
\begin{align}\label{GoodIdentity}
H^U H^U H^U H^U \mbox{vol}_{AAAA}\sim
\delta^{[U}_{A}\delta^{\vphantom{[}U}_A\delta^{\vphantom{[}U}_A\delta^{U]}_A
\;\mbox{vol}\quad .
\end{align}
Together with the algebraic properties of the Weyl tensors this results in
\begin{align}
\left.\delta S\right|_{\frac{\delta S_0}{\delta\varphi}=0}
= g\,\sum_k \frac{\alpha_k}{\beta_k}\;\sum_{m,n}\;\int_{M_d} tr\,
( C^k \star [C^m, \epsilon^n]_\star) \wedge \mbox{vol}
\end{align}
where $tr(...)$ means a projection to a singlet component, i.e. just
a total contraction of all indices (up to some factor), which is
unique. The extra coefficient $\beta_k^{-1}$ originates from using
Young symmetry properties to rearrange indices carried by $C^k$ and
from the fact that there can be a nontrivial normalization for
$tr(C^k\star C^k)$ that depends on $k$; $\mbox{vol}$ is a $d$-volume
form.

Later, we will argue that any higher-spin algebra admits a natural trace operation.
Then by making the choice $\alpha_k=\beta_k\,$, one can make appear traces
of commutators using
\begin{align}
tr\left(C^k\star[C^m,\epsilon^n]_\star + C^m \star
[C^k,\epsilon^n]_\star\right)\equiv - tr\left([C^m\star\epsilon^n,C^k]_\star
+[C^k\star\epsilon^n,C^m]_\star\right)\quad,
\end{align}
so that this expression identically vanishes by the definition of the trace.
Hence
\begin{align}
\delta S|_{\frac{\delta S_0}{\delta\varphi}=0}={-g} \sum_{k,n,m}
\int tr([C^k\star\epsilon^n,C^m]_\star)\wedge \mbox{vol} +
\mathcal{O}(\omega^3\epsilon)\quad.
\end{align}
Note that the action itself cannot be written in a trace-like form
$tr(R\wedge R)\,$, the latter action being topological. Fortunately,
when taken on free mass-shell, the gauge variation of the action
\emph{can} be written in terms of the trace on the algebra
$\mathfrak{g}\,$. Also note that one has to consider the
individual components $S^k$ in order to compute the factors
$\beta_k\,$, and it is important that $\beta_k$ depends on $k$ only
and not on $n$ and $m$ appearing in $\delta R^k$ via
(\ref{FVgaugeC}). The latter property is expected to hold true for
any action of type (\ref{FVaction}).

Let us also note that an action for a spin-one field
$S_{YM}=-\frac14\int tr (F_{\mu\nu})^2$ cannot be written in the
form (\ref{FVaction}) even if a spin-one field belongs to the
spectrum of the higher-spin algebra. Nevertheless, as was pointed out in
the original paper \cite{Fradkin:1986qy}, one can add $S_{YM}$ to
(\ref{FVaction}), where $F_{\mu\nu}$ is a projection of the full
curvature (\ref{FullCurvature}) to the spin-one sector, to get a
cubic action that includes vertices with spin-one. We will not
emphasize this subtlety in the sequel.

To conclude this section, once a higher-spin algebra is found it leads to an
action consistent up to the cubic  order, and one still has to find the free
quadratic action (\ref{FVaction}) and compute $\beta_k\,$.

%
\section{Gravitational interactions}
\label{sec:FVgravitational}
In this Section we would like to test gravitational interactions for the simplest
case of spin-$[k,1]$ gauge fields, i.e. we are interested in $[k,1]-[k,1]-[1,1]$
cubic vertices. This is a direct generalization of the case $k=2$ presented
in \cite{Boulanger:2011qt}.

We introduce the following set of one-form gauge fields
$\{e^a,\omega^{ab},e^{a[k]},\omega^{a[k+1]}\}$ where $\{e^a,\omega^{ab}\}$ are the
dynamical one-form gauge fields in the spin-2 sector. As recalled in Section \ref{sec:two},
the two fields $\{e^{a[k]},\omega^{a[k+1]}\}$ correspond to the one-forms needed to describe
an irreducible and unitary $[k,1]$-type gauge field in $\AdS$.

Quadratic corrections $R_1$ to curvatures (\ref{CurvatureA})-(\ref{CurvatureB})
are made by replacing background tetrad $h^a$ and Lorentz spin-connection $\varpi^{a,b}$
with $h^a+e^a$ and $\varpi^{a,b}+\omega^{a,b}$, respectively.
Quadratic contributions to the torsion and Riemann curvature are determined from
the most general Ansatz by requiring curvatures to be gauge invariant up to order $g$.
Denoting the total vielbein $\mathbf{e}^a=h^m+e^m\,$, the result is
\begin{eqnarray}
 T^{a} &=& D e^a + \mathbf{e}_b\,\omega^{ba} - g\,\omega^{ab[k]}\,e_{b[k]}\quad, \\
 R^{ab} &=& D \omega^{ab} + \omega^{a}{}_c\,\omega^{cb} - \Lambda \,\mathbf{e}^a \,\mathbf{e}^b
 - k\,\Lambda \, g \,e^{ac[k-1]}\,e^b{}_{c[k-1]} - g\, \omega^{ac[k]}\,\omega^b{}_{c[k]}\quad,  \\
 T^{a[k]} &=& D e^{a[k]} - \mathbf{e}_b\,\omega^{ba[k]} + \omega^{a}{}_b \,e^{ba[k-1]}\quad, \\
 R^{a[k+1]} &=& D \omega^{a[k+1]} +  \Lambda \mathbf{e}^a \,{e}^{a[k]} +
 \omega^{a}{}_b\,\omega^{ba[k]} \quad .
 \end{eqnarray}
The Yang--Mills-like gauge transformation are
\begin{eqnarray}
 \delta e^{a} &=&  {\rm d} \xi^a + {\xi}_b\,\omega^{ab} - {e}_b\,\xi^{ab} + g\,\eta^{ab[k]}\,e_{b[k]}
               - g\, \omega^{ab[k]}\,\eta_{b[k]} \quad,
     \\
 \delta \omega^{ab} &=& {\rm d} \xi^{ab} - \xi^{a}{}_c\,\omega^{cb} + \omega^{a}{}_{c}\,\xi^{cb}
              + \Lambda \,({\xi}^a \,e^b - e^a\,{\xi}^b)
              + k\,\Lambda \, g \,\eta^{ac[k-1]}\,e^b{}_{c[k-1]}
    \nonumber \\
              && - \;k\,\Lambda \, g \,e^{ac[k-1]}\,\eta^b{}_{c[k-1]}
              + g\, \eta^{ac[k]}\,\omega^b{}_{c[k]} - g\, \omega^{ac[k]}\,\eta^b{}_{c[k]}\quad,
      \\
  \delta e^{a[k]} &=& {\rm d} \eta^{a[k]} + {\xi}_b\,\omega^{ba[k]} - {e}_b\,\eta^{ba[k]}
              - \xi^{a}{}_b \, e^{ba[k-1]} + \omega^{a}{}_b \,\eta^{ba[k-1]} \quad,
      \\
  \delta \omega^{a[k+1]} &=& {\rm d} \eta^{a[k+1]} - \Lambda {\xi}^a \,e^{a[k]}
                + \Lambda {T}^a \,\eta^{a[k]} - \xi^{a}{}_b\,\omega^{ba[k]}
                +  \omega^{a}{}_b\,\eta^{ba[k]}  \quad ,
\end{eqnarray}
and accordingly, for the curvatures:
\begin{eqnarray}
 \delta T^{a} &=&  {\xi}_b\,R^{ab} - {T}_b\,\xi^{ab} + g\,\eta^{ab[k]}\,R_{b[k]}
               - g\,R^{ab[k]}\,\eta_{b[k]} \quad,
\label{deltaTa}\\
 \delta R^{ab} &=& - \xi^{a}{}_c\,R^{cb} + R^{a}{}_{c}\,\xi^{cb}
              + \Lambda \,({\xi}^a \,T^b - T^a\,{\xi}^b)
              + k\,\Lambda \, g \,\eta^{ac[k-1]}\,R^b{}_{c[k-1]}\nonumber \\
              && - \;k\,\Lambda \, g \,R^{ac[k-1]}\,\eta^b{}_{c[k-1]}
              + g\, \eta^{ac[k]}\,R^b{}_{c[k]} - g\, R^{ac[k]}\,\eta^b{}_{c[k]}\quad,
\label{deltaRab} \\
  \delta T^{a[k]} &=& {\xi}_b\,R^{ba[k]} - {T}_b\,\eta^{ba[k]}
              - \xi^{a}{}_b \,T^{ba[k-1]} + R^{a}{}_b \,\eta^{ba[k-1]} \quad,
\label{deltaTak} \\
  \delta R^{a[k+1]} &=& - \Lambda {\xi}^a \,T^{a[k]} +\Lambda {T}^a \,\eta^{a[k]}
                   - \xi^{a}{}_b\,R^{ba[k]} + R^{a}{}_b\,R^{ba[k]}  \quad
\label{deltaRak1}.
 \end{eqnarray}
The on-mass-shell linearized conditions for a free $[k,1]$-type fields read (\ref{tallomsA})
\begin{eqnarray}
&{\cal{T}}^{a[k]} ~=~ h_b h_b \; {\cal{ C}}^{a[k],b[2]} \quad,\qquad
{\cal {R}}^{a[k+1]}  ~=~ h_b h_b \; {\cal C}^{a[k+1],b[2]} \quad ,&
\label{onshellline}
\end{eqnarray}
while the spin-2 sector ($k=1$) gives the constraints
\begin{eqnarray}
{ \cal T}^a &=& 0\quad , \qquad { \cal R}^{a[2]} ~=~ h_b h_b \; {\cal W}^{aa,bb} \quad,
\label{onshellspin2}
\end{eqnarray}
where the linearized quantities are indicates by calligraphic symbols.

The \lorentz-tensors $\{ {\cal C}^{a[k],b[2]},{\cal C}^{a[k+1],b[2]}, {\cal W}^{a[2],b[2]}\}$
are irreducible tensors of symmetry type $[k,2],[k+1,2]$ and $[2,2]\,$, respectively.
\vspace*{.2cm}

We take the following Ansatz (\ref{HookActionA}) for the action, dropping the boundary term
\begin{eqnarray}
& S^{} ~=~ \frac{1}{2}\; \int  (
R^{uu}R^{vv} + a_1 \; T^{uua[k-2]} T\fud{vv}{a[k-2]}
 + a_2 \int R^{uua[k-1]}R\fud{vv}{a[k-1]})\wedge \mbox{vol}_{uuvv} \quad, &
 \label{dynactionAnsatz}
\end{eqnarray}
where it is understood that the quartic terms are neglected at this order in perturbation.
The variation of the above action can be evaluated using (\ref{deltaTa})--(\ref{deltaRak1}),
keeping only terms bilinear in the fields and linear in the gauge parameter.
In other words, after taking the gauge variation inside the action,
the curvatures are replaced by their linearized expressions that are then constrained
according to (\ref{onshellline}) and (\ref{onshellspin2}).

Denoting
\begin{align}
\boldsymbol{A}&=\int \mbox{vol} \; {\cal C}_{uu,vv} \,\eta\fud{v}{a[k-1]} \,{\cal W}^{va[k-1],uu}
\quad, \\
\boldsymbol{B}&=\int \mbox{vol} \;{\cal C}_{uu,vv} \,\eta\fud{v}{a[k]} \,{\cal W}^{va[k],uu}
\quad, \\
\boldsymbol{C}&=\int \mbox{vol} \;{\cal W}_{a[k],mn} \,\xi_c \,{\cal W}^{ca[k],mn}
\quad,
\end{align}
the Fradkin--Vasiliev consistency condition gives the following constraint on the free parameters
entering  the action $S[e^a,\omega^{ab},e^{a[k]},\omega^{a[k+1]}]$:
\begin{align}
 \left[ k \Lambda g - \frac{ a_1}{(k-1)}\; \right] \,\boldsymbol{A}
 + \left[ g - \frac{ a_2}{k}\;\right] \,\boldsymbol{B}
 + \left[ \frac{ a_1}{k(k-1)}\; - \frac{ a_2\Lambda }{k}\;\right] \,\boldsymbol{C} =0\quad.
\end{align}
This admits the solution
\begin{equation}\label{graviratio}
a_1~=~ g\, k(k-1)\,\Lambda  \quad, \qquad a_2~=~ g\,k \quad .
\end{equation}
Since the ratio $\frac{a_1}{a_2}\,$ is completely fixed by the consistency of the action
(\ref{dynactionAnsatz}), one must set  $\alpha=0$ in the action (\ref{HookActionA}).

Thus, a natural requirement to include cubic interactions with gravity,
gives us a less general Ansatz for action, in fact only one term is possible
in \ads-covariant language:
\begin{align}
S^{\ell}_0&=\frac{\alpha_{\ell}}{2}\,
\int R^{UUA[\ell-2]}\, \wedge R\fud{VV}{A[\ell-2]}\,
\wedge \mbox{vol}_{UUVV}
\label{TallHookAdsAction}
\end{align}
and the choice for compensator $V^{A}=|\Lambda|^{1/2}\delta^A_{d+1}$ gives exactly
the same value $(k-1)\Lambda$ for the ratio $\frac{a_1}{a_2}\,$, (\ref{graviratio}), that is required by
consistency of cubic interactions.

The second term possible, in which $R$ is contracted with $V$, turns
out to be forbidden. For the case of general mixed-symmetry fields,
it is still easy to see that all terms in the action that (i) are
contracted with a number of $V$; (ii) to which generalized Weyl
tensors contribute; must vanish. The reason is in that such terms
must cancel with analogous terms coming from the spin-two action,
i.e. from projections of $\omega \star \omega$ to $R^{A,B}$. In the
latter Weyl tensors contribute via $\star$-product that does not
depend on $V$. Hence the cancellation takes place only for the terms
with no $V$-contractions. Therefore, switching on gravitational
interactions reduces the freedom to add boundary terms.

\section{Cubic interactions: AdS-covariant formulation} \setcounter{equation}{0}
\label{sec:AdScov}

One of the simplest candidate higher-spin algebras one can consider is the
Clifford algebra \Cl{} for \ads, which can be realized as an algebra of
anticommuting symbol variables $\phi^A$ with the Clifford $\star$-product on functions of
$\phi^A$ instead of usual Grassmann multiplication.
The star product between two functions $F$ and $G$ of $\phi$ can be realized by
\begin{equation}
F(\phi) \star G(\phi) =   F(\phi)\;
\exp\left(\frac{\overleftarrow{\partial}_r}{\partial\phi^C} \;\;\eta^{CD}\,
\frac{\overrightarrow{{\partial}}_l}{\partial\phi^D}\right) \, G(\phi) =
F(\phi)  \exp[\overleftarrow{\partial}_r  \cdot
\overrightarrow{{\partial}}_l] \, G(\phi)
\quad .
\end{equation}
It features left ($\partial_l$) and right ($\partial_r$) derivatives with respect to the Grassmann-odd
variables $\{\phi^A\}\,$ and the arrows on the derivatives indicate on
which function they act.
This $\star$-product corresponds to Weyl ordering of the symbols $F(\phi)$ and $G(\phi)\,$.
In particular, it leads to
\begin{align}
&\{\phi^A,\phi^B\}_{\star}=2\eta^{AB} \quad, &&
\phi^A\star\phi^B=\phi^A\phi^B+\eta^{AB}\quad,
\\
&\phi^A\star F(\phi)=\left(\phi^A+
\frac{\overrightarrow{\pl}_l}{\pl\phi_A}\right)F(\phi) \quad,&&
F(\phi)\star
\phi^A=F(\phi)\left(\phi^A+\frac{\overleftarrow{\pl}_r}{\pl\phi_A}\right)\quad.
\end{align}
The trace is a projection to the singlet, $\phi$-independent, component, i.e. $tr(F(\phi))=F(0)$.

We would like to consider one-forms $W\fm{1}(\phi)$, whose expansion coefficients
$W^{A[h]}$ in
\begin{equation}
W\fm{1}(\phi) = \sum_h \frac{1}{h!}\; W^{A[h]}\fm{1}\phi_{A_1}...\phi_{A_h}
\end{equation}
are one-forms that take their values in totally antisymmetric tensor $\ads$-modules
and hence unify all spin-$\Ya{k,1}$ fields according to Section \ref{sec:two}.
However, one has to truncate the algebra to even polynomials in $\phi$ since the
gauge field $W\fm{1}^A\phi_A$ is known to describe a partially-massless
graviton \cite{Skvortsov:2006at}, which is nonunitary in anti-de Sitter,
and we see no other way to truncate away only the linear term
in $\phi\,$ from $W\fm{1}(\phi)\,$ while maintaining the associative algebra structure.
We thus restrict ourselves\footnote{This is not enough, however, because of dual descriptions
$W^{A[k+1]}$ and $W^{A[d-k]}$ that describe the same spin-$[k,1]$ field \cite{Skvortsov:2009zu}.
Thus, both $(k+1)$ and $(d-k)$ must be even, \emph{i.e.} the dimension $d$ must be odd.}
to the even subalgebra $\Cle{}$ of $\Cl{}\;$:
\begin{align}
W(\phi)=W(-\phi)\quad .
\end{align}
Therefore, the gauging of \Cle{} describes fields with spins\footnote{{There
is a natural limitation on the spin due to the spacetime dimension.
We remark that a free spin-$[2,1]$ field in $AdS_4\,$ is equivalent to a
Fierz--Pauli (or Fronsdal) spin-2 field in $AdS_4\,$.}}
$\Ya{2h-1,1}$, $h=1,\ldots , [(d+1)/2]\,$ and a spin-$1$.
The important point, in order to have the spin-2 field in the spectrum,
is that $\ads\subset\Cle{}\,$, the generators of $\ads$ being
\begin{align}
M_{AB}=\frac14[\phi_A,\phi_B]_\star\label{AdSGeneratorsA}\quad.
\end{align}

\paragraph{Component form.} Expanding the curvature
$R(\phi)=D_0 W(\phi)+g\,W(\phi)\wedge\star\, W(\phi)$ in terms of its Taylor
components $\frac{1}{(2h)!}\,R^{A[2h]}\phi_{A_1}...\phi_{A_{2h}}\,$,
one arrives at
\begin{align}
R^{A[2h_1]} = D_0 W^{A[2h_1]}+g\sum_{h_2,h_3} a^{h_1}_{h_2,h_3}\,
W^{A[\mathbf{h}-2h_3]C[\mathbf{h}-2h_1]}\wedge W\fud{A[\mathbf{h}-2h_2]}{C[\mathbf{h}-2h_1]}
\quad,
\end{align}
where
\begin{equation}
\mathbf{h}=h_1+h_2+h_3\,\quad .
\end{equation}
The Clifford $\star$-product gives the following expression for the
structure coefficients $a^{h_1}_{h_2,h_3}\,$:
\begin{align}
a^{h_1}_{h_2,h_3} = (-1)^{\mathbf{h}(\mathbf{h}+1)/2+h_1}
\frac{(2h_1)!}{\displaystyle\prod_{i=1,2,3}(\mathbf{h}-2h_i)!}\quad .
\end{align}
Note that if $W^{A[2h_i]}$ were taken to be a form of degree $q_i$ then one
would have
\begin{align*}
W\fm{q_2}^{A[\mathbf{h}-2h_3]C[\mathbf{h}-2h_1]}\wedge
W\fm{q_3}\fud{A[\mathbf{h}-2h_2]}{C[\mathbf{h}-2h_1]}
 =
(-)^{\mathbf{h}^2+q_2q_3}\,
W\fm{q_3}^{A[\mathbf{h}-2h_2]C[\mathbf{h}-2h_1]}\wedge
W\fm{q_2}\fud{A[\mathbf{h}-2h_3]}{C[\mathbf{h}-2h_1]}
\end{align*}
from which it follows that in certain cases there are accidental zeros in
the couplings.
If all fields are one-forms then it is easy to see that there is always a
nonvanishing contribution of $\{W^{A[2k]}, \wedge\star\, W^{A[2k]}\}$ to the
graviton curvature $R^{A[2]}$ and $\{W^{A[2k]}, \wedge\star\, W^{A[2]}\}$ to
$R^{A[2k]}$ itself, \emph{i.e.} all the fields interact with gravity and
contribute to the gravitational energy-momentum tensor as they should.
For one-forms in general all mutual ``two-to-one'' $2h_1-2h_2\rightarrow2h_3$ couplings
are nonzero
if and only if $\mathbf{h}$ is odd and all couplings vanish otherwise, when
$\mathbf{h}$ is even.

\paragraph{Cubic action.} Following the Fradkin--Vasiliev procedure, we replace
$R_0$ in (\ref{FVaction}) with the nonabelian Yang--Mills-like $R$'s, taking
our preliminary result (\ref{TallHookAdsAction}) into account.
The variation of
\begin{align}
S&=\sum_{h=1}^{[(d+1)/2]}\frac{\alpha_{h}}{2}\,
\int R^{UUA[2h-2]}\, \wedge R\fud{VV}{A[2h-2]}\,
\wedge \mbox{vol}_{UUVV}
\label{TallHookAdsAction2}
\end{align}
reads
\begin{align}
\delta S=\sum_{h_1,h_2,h_3}\alpha_{h_1}\delta S_{h_2,h_3}^{h_1}=\sum_{h_1}\alpha_{h_1}\sum_{h_2,h_3}\int R\fud{UU}{A[2h_1-2]} [R^{2h_2}_0, \xi^{2h_3}]_\star^{VVA[2h_1-2]} \,\mbox{vol}_{UUVV}\,\, .
\end{align}
Consider then the term $\delta S_{h_2,h_3}^{h_1}$ of the variation and use the on-mass-shell theorem (\ref{tallomsB}):
\begin{align}
\delta S_{h_2,h_3}^{h_1}&=\int H_M H_N C\fudu{UU}{A[2h_1-2]}{,MN}
\left[ H_P H_Q C^{2h_2,PQ}, \xi^{2h_3}\right]_\star^{VVA[2h_1-2]}\,
\mbox{vol}_{UUVV}\,\,,
\end{align}
where the notation $C^{2h_2,PQ}$ means that the indices $A[2h_2]$ of
$C^{A[2h_2],PQ}$ are involved in the $\star$-product.
The result of $C^{2h_2}\star \xi^{2h_3}$ has indices $VVA[2h_1-2]$ and carries
indices $PQ$ that are not affected by the $\star$-product.
Using (\ref{GoodIdentity}) and the fact that the Weyl tensors are traceless,
the indices $MN$ must be contracted with $VV$ only, whence $PQ$ have to be
contracted with $UU\:$
\begin{align}
\delta S_{h_2,h_3}^{h_1}&\sim\frac{2}{2h_1(2h_1-1)}\int
C_{A[2h_1],UU}\left[C^{2h_2,UU}, \xi^{2h_3}\right]_\star^{A[2h_1]}
\mbox{vol}\quad
\end{align}
where the prefactor comes from the simple identity
\begin{align}
C_{A[M-N]U[N],A[N]}\,\phi^{A}...\phi^A&=\frac{(M-N)!N!}{M!}\, C_{A[M],U[N]}\,
\phi^{A}...\phi^A
\end{align}
used in the case $M=2h_1\,$, $N=2\,$.
In order to pass from the total contraction of indices $C_{A[2h_1]}B^{A[2h_1]}$
to the trace $tr(C\star B)=a^{0}_{h_1,h_1}C_{A[2h_1]}B^{A[2h_1]}$ one needs to
compensate for $a^{0}_{h_1,h_1}=\frac{(-)^{h_1}}{(2h_1)!}\,$, which results in
\begin{align}
\delta S_{h_2,h_3}^{h_1}&\sim 2(-)^{h_1}(2h_1-2)!\int
tr\left(C\fud{2h_1,}{UU}\star\left[C^{2h_3,UU},\xi^{2h_2}\right]_\star\right)\,
\mbox{vol}\quad .
\end{align}
Therefore, taking into account the analysis of Section \ref{sec:FV},
the choice
\begin{equation}
\alpha_h=\frac{(-1)^h}{2(2h-2)!}
\label{AlphaCoef}
\end{equation}
is such that the variation of the cubic action,
evaluated on free shell and taken at order $g\,$, can be presented as a trace
of commutators, thereby solving the problem:
\begin{align*}
\delta S|_{\frac{\delta S_0}{\delta \varphi}=0}
=\sum_{h_1,h_2,h_3}\alpha_{h_1}\delta S_{h_2,h_3}^{h_1}|_{\frac{\delta
S_0}{\delta \varphi}=0} = 0 + {\cal O}(g^2)\quad.
\end{align*}
Introducing additional Clifford oscillators $\psi^U$, which are to be contracted
with the second group of indices of Weyl tensors,
$\frac1{2(2h_1)!}C^{A[2h_1],UU}\phi_A...\phi_A\psi_U\psi_U$, the variation of the
action becomes a trace on the $\Cl{}\times\Cl{}$. Actually, to cancel the
variation of the action, only the indices carried by gauge potentials must be
hidden into the trace, so one can keep the pair $UU$ in plain view.

Note that there are some terms in the variation that vanish by
themselves. These originate from cubic self-interactions. The
statement whose applicability spreads far beyond the case of type
$[k,1]$ fields\footnote{One can check that cubic self-interactions
are always consistent for all gauge fields described by one-form
gauge potentials.} is that the Fradkin--Vasiliev condition for cubic
self-interactions is identically satisfied. The reason is that there
is no representation carried by gauge parameter $\xi$ in the
symmetric tensor product of two \ads-modules corresponding to Weyl
tensor, e.g. for the case of gravity
\begin{align}\mbox{no singlet in\quad} \left(\parbox{20pt}
{\YoungBB}\bigotimes\parbox{20pt}
{\YoungBB}\right)_{\mbox{Sym}}\bigotimes\parbox{10pt}{\YoungAA}
\quad\Longleftrightarrow\quad C^{AA,BC}C\fdu{AA,B}{D}\xi_{C,D}\equiv0\;\,.
\end{align}

It is worth noting that $\alpha_h$ can be determined from a more simple
requirement that the quadratic action can be represented as a trace
on free shell. The same $\alpha_h$ (\ref{AlphaCoef}) ensures that
\begin{align}
\left.\rule{0pt}{12pt}S_0\right|_{\frac{\delta S_0}{\delta \varphi}=0}
=\int tr\left(C_{UU} \star C^{UU}\right)\, \mbox{vol}\quad .
\end{align}

Notice that the quantity $\alpha_h$ in (\ref{AlphaCoef})
is equal to $a^{0}_{h-1,h-1}$
up to a constant factor independent of $h\,$.
That suggests writing the cubic action in the form
\begin{align}
S&=\frac12\int tr (R^{UU} \wedge\star\, R\fud{VV}{})\,\wedge
\mbox{vol}_{UUVV}\label{TallHookAdsActionB}
\end{align}
where one still has to take off a pair of oscillators on each curvature to
undress two pairs of indices, these are not involved in taking $\star$-product.

\paragraph{Unitarity.}
When expressed in terms of metric-like field $\phi$, any frame-like
action $\int R\wedge R$ is proportional to $(-)^{spin}(\pl \phi)^2$,
where the sign factor depends linearly on spin. This factor is
irrelevant for type-$[2h-1,1]$ fields considered here as they differ
by an even number of indices. Thus, in order to make action
(\ref{TallHookAdsActionB}) unitary one has to insert imaginary unit
to the definition of the star product, which compensates for
unwanted $(-)^h$ in $\alpha_h$. Equivalently, one can inherit the
reality conditions from the $\mathfrak{osp}(1|2)$-algebra
\cite{Vasiliev:2004cm}.

\section{Developments and Discussion}
\label{sec:TallHooksDiscusssion}
%
\paragraph{Extension to higher degree forms.}
The previous analysis has been carried over in the case of one-form
gauge potentials taking their values in the even Clifford algebra
$\Cle$. At least two problems appear when trying to include forms of
higher degree. \hfill\phantom{a}
\begin{wrapfigure}{l}{6cm}
\includegraphics[height=2.20in, width=2.36in]{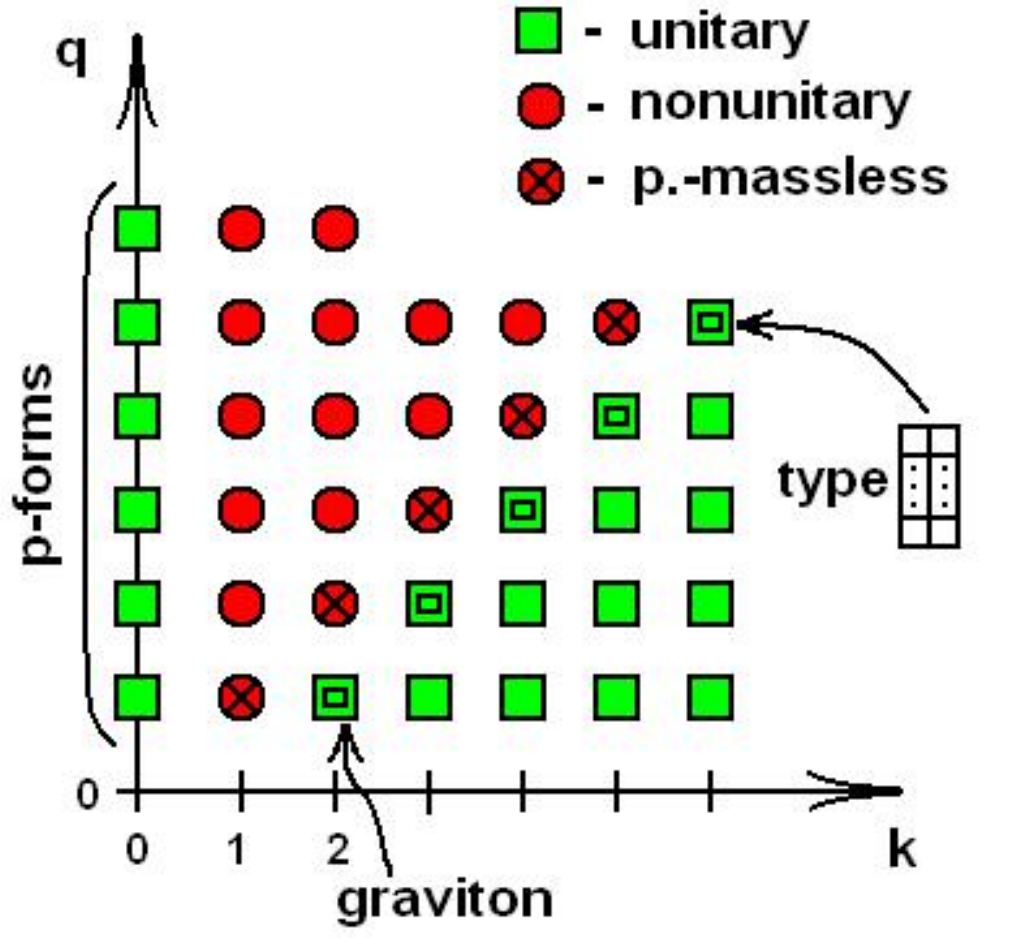}
\caption{Unitarity map.} \label{pict}
\end{wrapfigure}
Firstly, the number of forbidden values for $k$ in $W^{A[k]}\fm{q}$
grows with $q\,$. For example, among the one-forms only $W^A\fm{1}$
corresponding to a partially-massless graviton in $\AdS$ was
forbidden by unitarity. For two-forms, there are two such fields
$W^{A}\fm{2}$ and $W^{AA}\fm{2}$, \emph{etc}.
It is not evident how to truncate the algebra without breaking the
associativity in a way that the star-product of two unitary fields
gives no contribution to the sector of nonunitary fields. Secondly,
since higher degree forms cannot contribute to lower degree forms no
pairwise cancellation of terms is possible now. Generically, it is
difficult to imagine a theory for nonabelian $p$-forms with $p>1$
since the closure of the gauge algebra would require gauge
parameters with form degrees unbounded from above, which is hardly
compatible with a finite spacetime dimensionality.

\paragraph{Extension to higher orders.}
As is known \cite{Konshtein:1988yg}, the candidate higher-spin
algebra $\mathfrak{g}$ must satisfy the {\it admissibility
condition} in order for interaction of higher than cubic order to
exist. The {\it admissibility condition} demands that gauging
$\mathfrak{g}$, which describes certain field content, must match
the field content of some unitary representation of $\mathfrak{g}\,$.
For example, a little bit tautological though, gauging of \ads{}
itself leads to $e^a, \omega^{a,b}\in W^{A,B}$ that describes a free
spin-two field at the linearized level, the same time there exists a
unitary irreducible representation of \ads{} that is a spin-two
field.

The {\it admissibility condition} becomes highly nontrivial and
restrictive for genuine higher-spin algebras. In particular it was
found in \cite{Vasiliev:1988sa}, that certain higher-spin algebras
do not give rise to consistent theories beyond the cubic
approximation, these defective higher-spin algebras were shown
\cite{Konshtein:1988yg, Konstein:1989ij} not to meet the {\it
admissibility condition}, i.e. not to have any unitary representation
with spectrum giving by gauging thereof.

Therefore, the cubic approximation is insensitive to the {\it
admissibility condition}. Nevertheless, we may argue that $\Cle$
does not satisfy the {\it admissibility condition} because its
gauging leads to a too small spectrum, which is much smaller than
the one resulting from tensoring the minimal representations
$|Di\rangle$, $|Rac\rangle$ corresponding to conformal scalar and
spinor \cite{Vasiliev:2004cm}. It is still interesting to see if the
action closes at the quartic level as it happens for pure gravity.

\paragraph{Universal enveloping realization.}
\footnote{E.S. is grateful to Per Sundell for many valuable
discussions on \cite{Iazeolla:2008ix} and to M.A.Vasiliev for sharing
his draft.} In the spirit of the approach used in
\cite{Iazeolla:2008ix}, let us put some remarks on the universal
enveloping algebra $U(\mathfrak{h})$, $\mathfrak{h}\cong\ads\,$, and
on the explicit realization of $\mathfrak{g}=\Cle$ as a quotient of
$U(\mathfrak{h})\,$. Consider $U(\mathfrak{h})$, generated by
$M_{AB}$ modulo relations ($\star$ denotes the product in
$U(\mathfrak{h})$)
\begin{align}
[M_{AB},M_{CD}]_{\star}&=M_{AD}\eta_{BC}-M_{BD}\eta_{AC}-
M_{AC}\eta_{BD}+M_{BC}\eta_{AD}\quad.
\end{align}

It is useful to write down the decomposition of the first several levels of
$U(\mathfrak{h})$ in terms of the standard adjoint action of $\mathfrak{h}\,$,
which by the Poincare--Birkhoff--Witt theorem is equivalent to computing
symmetric products of $M_{AB}\sim\parbox{10pt}{\YoungAA}\,$,

\begin{align}\label{SOspectrum}
\left. U(\mathfrak{h})\right|_{\mathfrak{h}}
&\cong \underbrace{\bullet}_{0}\oplus\underbrace{\left(\;\parbox{10pt}
{\YoungAA}\;\right)}_1\oplus
\underbrace{\left(\parbox{20pt}{\YoungBB}\oplus\parbox{10pt}
{\YoungAAAA}\oplus\parbox{20pt}{\YoungB}\oplus\bullet\right)}_2
\oplus
\underbrace{\left(\;\parbox{30pt}{\YoungCC}\oplus\ldots\;\right)}_3\oplus
\ldots
\end{align}
where the singlet $\bullet$ at the level zero is the identity of
$U(\mathfrak{h})$ and another one at the level two is the quadratic
Casimir operator $C_2=-\frac12M_{AB}\star M^{AB}\,$. All singlets in
$U(\mathfrak{h})$ are by definition certain functions of the Casimir
operators $C_{2i}$, $i=1,..,N\,$, $N=[\frac{d+1}2]$. The center of
$U(\mathfrak{h})$ is a free field $K[C_2,...,C_{2N}]$ in $N$
variables.

According to \cite{Iazeolla:2008ix} a higher-spin algebra $\mathfrak{g}$ can be
constructed as a quotient algebra of $U(\mathfrak{h})$ over a given
two-sided ideal. The ideal corresponding to the Vasiliev $sp(2)$ higher-spin
algebra \cite{Vasiliev:2003ev}, whose gauging describes all totally-symmetric
massless fields, is generated by two $\mathfrak{h}$-covariant elements,
\begin{align}
&I\cong
U(\mathfrak{h})\star\left(\;\parbox{10pt}{\YoungAAAA}\oplus\parbox{20pt}
{\YoungB}\;\right)\star U(\mathfrak{h})\quad,\end{align}
where
\begin{align}
&\parbox{10pt}{\YoungAAAA}=M_{[AB}\star M_{CD]}, &&
\parbox{20pt}{\YoungB}=M\fdu{A}{C}\star M\fdu{AC}{}
-\frac{2}{(d+1)}\eta_{AA}C_2\,.
\end{align}
Roughly speaking, to quotient by $I$ means that all diagrams with
more than two rows as well as all the elements $M\star...\star M$
where at least two $\mathfrak{h}$-indices are contracted must be set
to zero. The resulting $\mathfrak{h}$-adjoint spectrum of
$U(\mathfrak{h})/I$ is given by all rectangular two-row diagrams,
\begin{align}
\left.\mathfrak{g}=U(\mathfrak{h})/I\right|_{\mathfrak{h}}\cong\bullet\oplus\parbox{10pt}
{\YoungAA}\oplus\parbox{20pt}{\YoungBB}\oplus\parbox{30pt}{\YoungCC}
\oplus\ldots
\end{align}
The salient feature of the ideal $I$ is that besides sorting out
`unwanted' diagrams it also restricts all Casimirs, $C_{2i}$, to
particular values $\mu_{2i}\,$. When inside $U(\mathfrak{h})$, to
determine the values of $C_{2i}$ one \cite{Iazeolla:2008ix} has to
verify the consistency of the ideal by multiplying its elements and
inspecting if the result belongs to the ideal too. This procedure
leads for example to relations of the type $(C_{2i}-\mu_{2i})\star
M_{AB}\sim0$, of which the nontrivial solution is
$C_{2i}-\mu_{2i}\sim0$. Note that if $C_{2i}$ were free it would
lead to a degeneracy of the spectrum due to the center $K[C_2,...]$
of $U(\mathfrak{h})/I$.

As was noticed in \cite{Vasiliev:1999ba,Iazeolla:2008ix}, the ideal
$I$ is in fact the annihilator, $\mbox{Ann}(|Rac\rangle)\,$, of the
remarkable Dirac scalar singleton representation $|Rac\rangle$ of
$\mathfrak{h}\,$, which fixes all $C_{2i}$ accordingly. Therefore,
quite generally one may think of any higher-spin algebra
$\mathfrak{g}$ as the universal enveloping algebra $U(\mathfrak{h})$
of $\mathfrak{h}$ evaluated in some $\mathfrak{h}$-module, say $V$,
\begin{align}
\mathfrak{g}=\left.\rule{0pt}{12pt}U(\mathfrak{h})\right|_{V}\sim
End(V)\sim V^*\otimes V\,.
\end{align}

The scalar $|Rac\rangle$ and spinor $|Di\rangle$
singletons are however very distinguished representations.
We see no natural way to generalize $|Rac\rangle$, $|Di\rangle$ to some other
representation, say $V$, such that
$\mathfrak{g}=U(\mathfrak{h})/\mbox{Ann}(V)$ would contain
$\mathfrak{h}$-modules $A_i$, $\mathfrak{g}=\oplus_i A_i$ in its
$\mathfrak{h}$-adjoint decomposition that by means of $A_i$-valued
generalized connection $\omega\fm{q}^{A_i}$ would describe unitary
mixed-symmetry fields for some $q\,$. As noted in
\cite{Boulanger:2008kw,Bekaert:2009fg} higher-spin singletons should
give examples of higher-spin algebras, whose gauging leads to
certain multiplets of mixed-symmetry fields. However, these exist
only in odd dimensions and we do not expect that consistent theories
with mixed-symmetry fields are confined to odd dimensions. Moreover,
the tensor product of two higher-spin singletons with high enough spins
does not contain a graviton. As the most simple example, using
\cite{Sezgin:2001zs} one can evaluate the product
$|j_1,0\rangle\otimes|0,j_2\rangle$ of two $\mathfrak{so}(4,2)$
higher-spin singletons with spins $j_1$ and $j_2$, which are also
called doubletons, \cite{Gunaydin:1998km},
\begin{align}
|j_1,0\rangle\otimes|0,j_2\rangle=\bigoplus_k\left[
\substack{\displaystyle \mbox{massless unitary}\\ \displaystyle
\mbox{mixed-symmetry}\\ \displaystyle\mbox{fields in }
AdS_5}\right]\mbox{ of type }
\parbox{60pt}{\RectBRow{6}{4}{$j_1+j_2+k$}{$j_1-j_2$}},\end{align}
The sign of $(j_1-j_2)$ distinguishes between selfdual and
anti-selfdual fields. This result reduces at $j_1=j_2=0$ to the
Flato-Fronsdal-type theorem \cite{Flato:1978qz} of
\cite{Sezgin:2001zs} that the product of two scalar singletons
decomposes into a sum over all totally symmetric bosonic higher-spin
fields, see also \cite{Vasiliev:2004cm, Dolan:2005wy}.

Thus the ability of higher-spin singletons to describe a world with
gravity and mixed-symmetry fields is very restricted. Therefore, as
we have no candidates for the annihilator, we would like to define
$I$ directly by specifying which diagrams are `unwanted'.

The adjoint spectrum of $\mathfrak{g}=\Cle$ consists of all $[2k]$ types with
multiplicity one,
\begin{align}
\mathfrak{g}&=\underbrace{\bullet}_{0}\oplus\underbrace{\left(\parbox{10pt}
{\YoungAA}\right)}_1\oplus
\underbrace{\left(\parbox{10pt}{\YoungAAAA}\right)}_2\oplus...
\end{align}
This suggests the ideal be generated by
\begin{align}
\parbox{20pt}{\YoungBB}\oplus\parbox{20pt}{\YoungB}\sim I_{AA,BB}=M_{AB}\star
M_{AB}+\frac{2C_2}{d(d+1)}(\eta_{AA}\eta_{BB}-\eta_{AB}\eta_{AB})\sim0
\end{align}
Indeed, in verifying the compatibility condition $M^{AB}\star I_{AA,BB}\sim0$,
which can be done by using the following relation, which holds true modulo terms
proportional to $I\fdu{AA,B}{B}$,
\begin{align}
M\fdu{A}{C}\star M\fdu{BC}{}\sim \frac{-2C_2}{d+1}\eta_{AB}-\frac{d-1}{2}M_{AB}
\end{align}
one finds that the Casimir must be a fixed number
$C_2=\frac{d(d+1)}{8}$,
\begin{align}
0\sim M^{AB}\star I_{AA,BB}\sim \left(C_2-\frac{d(d+1)}{8}\right)\star
M_{AB}\sim0\quad,
\end{align}
which is exactly $C_2$ computed in the representation
(\ref{AdSGeneratorsA}) and is, as expected, equal to the Casimir of
the spinor module. Therefore\footnote{Unfortunately, it is very
complicated to inspect all relations that come from
$U(\mathfrak{h})\star I\star U(\mathfrak{h})$ for some $I$, in
particular, to find out if there are some additional relations at
higher levels, e.g. at the level $[(d+1)/2]$, which restricts $d$ to
be odd.},
\begin{align}
\mathfrak{g}&=U(\mathfrak{h})/\left(\rule{0pt}{10pt}U(\mathfrak{h})\star
I_{AA,BB}\star U(\mathfrak{h}) \right)\quad.
\end{align}
The analog of $|Rac\rangle$ representation for $\Cle$ is the spinor
representation,
which is finite dimensional in accordance with finiteness of $\Cle$ (the symmetry
algebra of a field equation, e.g. conformal scalar, must be an infinite
dimensional algebra as it contains arbitrary powers of translation generators).

In general, we see from (\ref{SOspectrum}) that any reasonable ideal $I$ must
take away $\parbox{20pt}{\YoungB}$ at the least,
\begin{align}
& I_1=\left(U(\mathfrak{h})\star
\parbox{20pt}{\YoungB} \star U(\mathfrak{h})\right)
\end{align}
as the generalized connection $W^{AB}\fm{q}=W^{BA}\fm{q}$ describes
a nonunitary theory for any $q$, \cite{Alkalaev:2003qv,
Skvortsov:2009zu}. This requirement already forces higher Casimirs,
$C_{2i}$, $i=2,...$,
\begin{align}
C_{2k}&=\frac12 M\fdu{A}{B}\star M\fdu{B}{C}\star...\star M\fdu{U}{A}
\end{align}
to be certain functions $F_{2i}(C_2)$ of the quadratic one
\begin{align*}
F_{2i}(X)&=\frac{X\left((\lambda_+)^{2i-1}-(\lambda_-)^{2i-1}\right)}
{\lambda_+-\lambda_-}, &\lambda_\pm&=\frac{d-1}4\left(1\pm\sqrt{1-\frac{32}{(d+1)(d-1)^2}X}\right)\quad,\\
F_2(X)&=X, & F_4(X)&=\frac{X}{4}\left(\frac{8X}{d+1}+(d-1)^2\right)\quad.
\end{align*}

Therefore, the only `degree of freedom' left if is due to $C_2\,$.
In the spirit of Feigin's $\mathfrak{gl}(\lambda)$, \cite{Feigin},
which is equivalent to the deformed oscillators of
\cite{Vasiliev:1989re, Prokushkin:1998bq}, we can quotient further
$(C_2-\nu)$ and define a one parameter family of higher-spin
algebras
\begin{align}
hs(\nu)=U(\mathfrak{h})/\left(U(\mathfrak{h})\star\{
\parbox{20pt}{\YoungB}\oplus(C_2-\nu)\} \star
U(\mathfrak{h})\right)\,.
\end{align}
At certain values of $\nu$ the algebra $hs(\nu)$ acquires an
ideal that can be quotient out, giving a smaller algebra. The value
of $C_2$ would be fixed by choosing one more `unwanted' diagram in
the $\mathfrak{h}$-adjoint spectrum of $U(\mathfrak{h})\,$. The
$\mathfrak{h}$-adjoint decomposition of $hs(\nu)$ is easy to
describe as
\begin{align}
\left.hs(\nu)\right|_{\mathfrak{h}}&=\left\{\mbox{diagrams made out
of }
\parbox{10pt}{\YoungCcA}\right\}\quad,
\end{align}
where the new elementary cell $\parbox{10pt}{\YoungCcA}$ denotes
$\parbox{10pt}{\YoungAA}$\,. Note that by inspecting generalized
connections of \cite{Alkalaev:2003qv} (that are allowed by
unitarity) we conclude that gauging of $hs(\nu)$ can describe
unitary fields. The additional ideal $I_2$ leading to the Vasiliev
$sp(2)$ higher-spin algebra, which removes the degeneracy due to
$C_2$, is generated by $\parbox{10pt}{\YoungCcAA}$.\footnote{Note
that by $\parbox{10pt}{\YoungCcAA}$ we do not mean the antisymmetric
tensor product of $\parbox{10pt}{\YoungAA}$ with itself, but simply
the Young diagram with four cells in the same column.} The ideal
$I_2$ leading to $\Cle$ is $\parbox{20pt}{\YoungCcB}$. The ideal
$I_2$ leading to the $\phi$-even Vasiliev $osp(1|2)$ higher-spin
algebra \cite{Vasiliev:2004cm} is $\parbox{20pt}{\YoungCcBB}\,$. Let
us note that there is no room here for the hypothetical algebra
whose spectrum was suggested in \cite{Sorokin:2008tf} in the context
of reducible multiplets of totally-symmetric fields, to read
\begin{align}
\underbrace{\bullet}_0\oplus\underbrace{\parbox{10pt}{\YoungCcA}}_1
\oplus\underbrace{\left(\rule{0pt}{10pt}\parbox{20pt}{\YoungCcB}\oplus\bullet\right)}_2
\oplus\underbrace{\left(\rule{0pt}{10pt}\parbox{30pt}{\YoungCcC}\oplus\parbox{10pt}{\YoungCcA}\right)}_3
\oplus\underbrace{\left(\rule{0pt}{10pt}\parbox{40pt}{\YoungCcD}\oplus\parbox{20pt}{\YoungCcB}\oplus\bullet\right)}_4
\oplus...\quad.
\end{align}
The above spectrum may result simply from extending the field of scalars.

Note that factoring out any nontrivial component of $U(\mathfrak{h})$ at the level-$(k+1)$ expresses higher Casimirs $C_{2k+2}$,... in terms of lower ones $C_2,...,C_{2k}$.
For example, which is relevant below, $\parbox{10pt}{\YoungCcAA}\sim0$ leads to
\begin{align}
&M_{[UU}\star M_{UU]}\sim0 && \Longrightarrow && M_{[UU}\star ... \star M_{UU]}\star M_{[VV}\star ... \star M_{VV]}\eta^{UV}...\eta^{UV}\sim0
\end{align}
that allows to express any $C_{2k}$, $k>1$ as a function $G_{2k}(C_2)$ of $C_2$, e.g. $G_4$ is
\begin{align}
&C_4=G_4(C_2)=(C_2)^2+\frac{(d-2)(d-1)}{2}C_2\,.
\end{align}
It is interesting to find out if there exist ideals generated by
more than two diagrams, the natural restrictions coming from the
condition that any $k$-generated ideal lies in the intersection of
$k$ algebraic functions of Casimirs. For example, the scalar
singleton point corresponding to the Vasiliev higher-spin algebra is
the unique intersection of $F_{2i}(X)=G_{2i}(X)$.

If one factors out only $\parbox{10pt}{\YoungCcAA}$ without
factoring out $\parbox{20pt}{\YoungB}$, the resulting spectrum is
rich enough to describe all massive totally-symmetric fields in the
spirit of \cite{Ponomarev:2010st}. The degeneracy due to $C_2$
enlarges the spectrum with St\"uckelberg companions. Flow with
respect to $C_2$ should pass all critical points where massive
fields decompose into partially-massless fields plus massive fields
of lower spin.

It seems natural that in addition to $I_1$ one may pick any
$\parbox{10pt}{\YoungCcA}$-diagram, say $\Yy$, from the spectrum of
$A_1$ and build the quotient algebra whose spectrum does not contain
the $\parbox{10pt}{\YoungCcA}$-diagrams for which $\Yy$ is a
subdiagram. The value of $\nu$ is to be determined by
consistency of $I_\Yy=U(\mathfrak{h})\star\Yy\star U(\mathfrak{h})$
with $I_1\,$.

As was mentioned in the introduction, $\Cle$ is just the simplest higher-spin algebra in the hierarchy, which
now can be depicted as
\begin{align*} &\ads\quad\rightarrow\overbrace{\quad \Cle\quad \rightarrow \quad
hs(\nu_i)\quad}^{\mbox{finite-dimensional}}
\rightarrow\overbrace{\quad osp(1|2)_e\quad\rightarrow\quad
hs(\nu) \quad}^{\mbox{infinite-dimensional}} \\
&
\phantom{\ads}\rotatebox{-4}{\bep(0,0)\put(3,15){\vector(1,0){160}}\eep}
\phantom{\quad\rightarrow\underbrace{\quad \Cle\quad \rightarrow
\quad hs(\nu_i)\quad}\rightarrow}\quad
sp(2)\bep(0,0)\put(3,8){$\nearrow$}\eep\end{align*} where
$\rightarrow$ means succession of gauging, i.e. the spectrum of
fields resulting from gauging one algebra belongs to the gauging of
the next one. Together with the Clifford algebra one can construct
other finite-dimensional algebras that correspond to $hs(\nu)$ at
certain $\nu_i$ of finite-dimensional modules as well as
infinite-dimensional algebras that correspond to $hs(\nu)$ at
generic $\nu$ or specific $\nu$'s of infinite-dimensional modules.

That the generators of a higher-spin algebra $\mathfrak{g}$ obtained
from $U(\mathfrak{h})$ have only even ranks (number of indices of
$\mathfrak{h}$) seems to be a drawback of $U(\mathfrak{h})$ as of
any two mixed-symmetry fields whose ranks differ by one index one
can be realized as a part of such a $\mathfrak{g}$ while another one
cannot. The possible way out is shown by the
$\mathfrak{osp}(1|2)$-algebra \cite{Vasiliev:2004cm}, that is to
extend $U(\mathfrak{h})$ with Clifford algebra. However, in this way
it is still not possible to have an algebra whose gauging leads, for
example, to two-row mixed-symmetry fields only, as Clifford algebra
brings a one-column Young diagram of height up to $d+1$ that is
attached at the bottom.

The above consideration within the universal enveloping algebra is
not the most general one. The $\star$-product of $U(\mathfrak{h})$
prescribes a particular way of contracting indices when expanding
expressions like $\omega\star\omega$ in components.

For example, for the Vasiliev $\mathfrak{sp}(2)$ algebra
$\star$-product is induced by $\exp{\left(s\right)}$, where $s=
t^{\alpha\beta}C_{\alpha\beta}$,
$t^{\alpha\beta}=\frac{\overleftarrow{\pl}}{\pl
Y^A_\alpha}\frac{\overrightarrow{\pl}}{\pl Y^B_\beta}\eta^{AB}$.
This star product is in fact $\mathfrak{sp}(2d+2)$ invariant as it
makes use of $\eta^{AB}C_{\alpha\beta}$. Various other
$\mathfrak{sp}(2)\oplus\mathfrak{so}(d-1,2)\subset\mathfrak{sp}(2d+2)$
invariants built with $t^{\alpha\beta}$ can be used to contract
indices. The field of $\mathfrak{sp}(2)$ invariants for
$t^{\alpha\beta}$ is generated by $s$ and $p=\det{t^{\alpha\beta}}$.
An arbitrary nontrivial monomial $s^k p^n$ gives rise to consistent
cubic interactions at least at the cubic level
\cite{Boulanger:2011xx}, the $\star$-product being different from
the one dictated by $U(\mathfrak{h})$. This is to be compared with
the general formalism of cubic interactions developed in the
very recent paper by Vasiliev \cite{Vasilev:2011xf}.

In general one is led to study the full tensor algebra of $\mathfrak{h}$.
The Grothendieck ring of tensor category of $\mathfrak{h}$-modules is an
associative ring whose basis $e_\mu$ is enumerated by all finite-dimensional $\mathfrak{h}$-modules
$\mu$
and the structure constants are defined by the decomposition of
$\mu\otimes\nu$ into irreducibles.
Inevitably any higher-spin algebra corresponds to a subring of the Grothendieck
ring, the additional requirement being that $\mathfrak{h}$ itself as its adjoint
module must belong to the higher-spin algebra. Finding an appropriate truncation
of the Grothendieck ring does not solve the problem yet, as one has to choose a
particular way to contract indices, which we believe can be done by classifying
invariants, like $s$ and $p$, corresponding to the truncation.

Then, the trace operation $tr$ can be defined as a projection to the singlet
component since it is unique. $tr(AB-BA)=0$ holds automatically because if $A$ is
isomorphic to $B$ as $\mathfrak{h}$ modules then there is no singlet component in
$A\wedge B$, otherwise $A\otimes B$ does not contain a singlet component either.
Studying subrings of the Grothendieck ring might be useful for finding
higher-spin algebras whose gauging leads to a desired spectrum of fields.
We hope to come back to this issue in a future work.

\section{Conclusion}
\label{sec:Conclusion}

We have proven that the Clifford algebra correctly produces not only the
structure but all the coefficients that are required for type $[2k-1,1]$ fields,
to have consistent interactions with gravity and themselves at the cubic order.
It is instructive to investigate if the same action can be made consistent up to
the quartic order as it does for the sector of pure gravity and if not where the obstructions come from.

We have seen that the Fradkin-Vasiliev approach is a powerful
machinery, which allows one to construct cubic vertices for a
multiplet of higher-spin fields once the candidate higher-spin
algebra $\mathfrak{g}$ is known.  The spectrum of fields is given by
the adjoint decomposition of $\mathfrak{g}$ with respect to the
anti-de Sitter algebra, leading to a number of generalized
connections of \ads. One requires quadratic actions of type $\int
R_0\wedge R_0$ as an input. The Fradkin-Vasiliev recipe is to
replace the linearized curvatures $R_0$ with the nonlinear ones $R$
that are dictated by the algebra and adjust coefficients in front of
individual actions to push the gauge invariance to the next
nontrivial order. The procedure determines all coupling constants in
front of different cubic vertices in terms of just one constant.

Once the quadratic actions built with curvatures $R_0$ for
generalized connections are known one needs to look at those parts
of the action to which the generalized Weyl tensors $C$ contribute
via the on-mass-shell theorem. The coefficients are then determined
by requiring $R_0\wedge R_0$ terms to have the form $tr(C_{UU}\star
C^{UU})$ on-mass-shell, which is reminiscent of Yang-Mills'
$tr(F_{\mu\nu}\star F^{\mu\nu})$. This gives a combinatoric factor
originating from using Young symmetry properties and the
normalization of the $\mathfrak{g}$-trace; the computation can be
done for a free action and is quite simple.

Following \cite{Iazeolla:2008ix} we believe that the universal
enveloping algebra of $\mathfrak{so}(d-1,2)$ is a natural framework
for description of higher-spin fields and we have treated $\Cle$
from this point of view. We have also discussed some general features
of embedding a higher-spin algebra into $U(\mathfrak{so}(d-1,2))$,
which results in that any reasonable higher-spin algebra built from
$U(\mathfrak{so}(d-1,2))$ should correspond to $hs(\nu)$ at a
particular $\nu\,$.

\newpage
\section{Acknowledgements}
\label{sec:Aknowledgements}
E.S. would like to thank E. Feigin, R. Metsaev, Yu. Zinoviev, M.A.
Grigoriev, O.V. Shaynkman, P. Sundell, K.B. Alkalaev, A. Campoleoni
and M. A. Vasiliev for many valuable discussions. N.B. thanks F.
Buisseret, P. P. Cook, A. Campoleoni, P. Sundell and Yu. Zinoviev
for discussions. E.S. acknowledges communications with M. A.
Vasiliev and thanks the service de M\'ecanique et Gravitation at
UMONS for hospitality. The work of E.S. was supported in parts by
RFBR grant No.11-02-00814 and President grant No.5638. The work of
N.B. was supported in parts by an ARC contract No.
AUWB-2010-10/15-UMONS-1.


\providecommand{\href}[2]{#2}\begingroup\raggedright\endgroup

\end{document}